\documentclass[]{pasj01}
\draft

\Received{2019/04/15}
\Accepted{2019/08/00}

\begin{document}

\title{Overall variation of  the H$_2$O masers around W~Hydrae in 28 years } 

\author{Hiroshi  \textsc{Imai}\altaffilmark{1,2}, Akiharu \textsc{Nakagawa}\altaffilmark{1}, and Hiroshi \textsc{Takaba}\altaffilmark{3}}

\altaffiltext{1}{Amanogawa Galaxy Astronomy Research Center, Graduate School of Science and Engineering, Kagoshima University,  \\
1-21-35 Korimoto, Kagoshima 890-0065}
 \email{hiroimai@sci.kagoshima-u.ac.jp}

\altaffiltext{2}{Center for General Education, Institute for Comprehensive Education, Kagoshima University,  \\ 
1-21-30 Korimoto, Kagoshima 890-0065}

\altaffiltext{3}{Faculty of Engineering, Gifu University, 1-1 Yanagido, Gifu, Gifu 501-1193, Japan}


\maketitle


\begin{abstract}
In this paper, we present the distribution of H$_2$O masers associated with a semi-regular variable star W Hydrae (W~Hya). We have collected the radio interferometric data of the maser distribution taken with the Very Large Array (VLA), the Kashima--Nobeyama InterFErometer (KNIFE), Multi-Element Radio Link Network (MERLIN), the VLBI Exploration of Radio Astrometry (VERA), and the combined array of Korean VLBI Network (KVN) and VERA (KaVA) in order to trace the maser distribution variation in two decades. Even though differences in the sensitivities and angular resolutions of the interferometric observations should be taken into account, we attempt to find possible correlation of the maser distribution with the stellar light curve. Our failure in measurement of the annual parallax of the masers with VERA is likely caused by the properties of the maser features, which have been spatially resolved by the synthesized beam and survived for only half year or shorter. No dependence of the maser spot flux density on its size is found in the KNIFE data, suggesting that maser spot size is determined by the physical boundary as expected for a clump affected by outward propagation of a stellar pulsation shock wave rather than the (spherical) geometry of maser beaming in the maser gas clump.  
\end{abstract}


\section{Introduction}

Cosmic H$_2$O masers are mainly associated with circumstellar envelopes (CSEs) of long-period ($>$100~d) variable stars in the asymptotic giant branch (AGB) or post-AGB phases, star-forming regions in relatively early stages of newly-born stars ($t<10^5$~yr, e.g., \cite{gen77}), and circumnuclear disks around super-massive black holes in external galaxies. These H$_2$O masers have been investigated as important probes of the spatio-kinematics of the regions around the host objects in detail through radio interferometric observations (see review of e.g. \cite{gra12}). Some of the maser sources have their trigonometric parallax distances determined in high accuracy astrometry \citep{rei14}. 

The excitation conditions of circumstellar H$_2$O masers should be a big issue in these applications and have been estimated from apparent distributions of gas clumps emitting the maser emission, each of which corresponds to a ``maser feature".  A series of successful trigonometric observations of H$_2$O masers (e.g., \cite{nak14}) suggest that motions of some maser features are stable over one year without strong and rapid accelerations/decelerations although their fraction is quite limited. On the other hand, propagation of shock waves induced by stellar pulsation may cause rapid accelerations unless the maser features move together with the shocks \citep{ima03,ric12}. The stability of maser features would be investigated through intensive interferometric monitoring observations as conducted for circumstellar SiO maser sources (e.g., \cite{gon14}). On the longer timescale over stellar pulsation cycles, rapid and periodic change in the maser distribution is expected due to periodic change in the physical condition for maser excitation \citep{ric10}. In this case, apparent rapid inward/outward propagation of maser emission region can be seen.  \citet{bow94} pointed out for the first time possibility of periodic variation in the maser distribution around RR Aql. However, the proposed periodicity is still obscure due to too sparse monitoring of the maser distribution. 
 
In this paper, we present the long term performance of H$_2$O masers around a semi-regular variable star (SRa) W Hydrae (W~Hya) found in a variety of interferometric data taken for two decades. \citet{gaia18} reported a trigonometric parallax of W~Hya to be $\pi=6.09\pm 0.82$~mas, corresponding to a distance of $D=164^{+25}_{-35}$~pc. However, this parallax is much smaller than the apparent stellar size ($\sim$30~mas) and the position of the referenced stellar photo-center may be significantly affected by the inhomogeneity of the stellar surface \citep{zha11}. In spite of its proximity to the Sun, its distance still has a large uncertainty, from 65~pc to 100~pc (e.g., \cite{olo02, vle03}). For a similar reason, the individual velocity components of maser emission (maser spots) around W~Hya have been often spatially resolved out in very long baseline interferometry (VLBI) observations because they are apparently larger than the synthesized beams ($\leq$10 mas, e.g. \cite{ima97}). This paper shows the distribution of only {\it compact} H$_2$O maser spots/features around W~Hya and describes the difficulty in determining their trigonometric parallax. Nevertheless, the presented interferometric and astrometric observations will provide some clue of the periodicity of the maser distribution and the physical condition of the maser regions.  



\section{Observations and data reduction}

This paper collects the results of interferometric observations of H$_2$O masers around W~Hya since 1985. We obtained the data taken with the Very Large Array (VLA), the Kashima--Nobeyama InterFErometer (KNIFE), Multi-Element Radio Link Network (MERLIN), the VLBI Exploration of Radio Astrometry (VERA), and the combined array of Korean VLBI Network (KVN) and VERA (KaVA). Table \ref{tab:observations} summarizes those interferometric observations, with the light curve phases estimated as described in Appendix \ref{sec:light-curve}. The references of the previous VLA and MERLIN data are given there. The numerical VLA data were unavailable from \citet{rei90}. In the next subsections, we describe our own VLBI observations with KNIFE, VERA, and KaVA. 

\subsection{Imaging observation with KNIFE}
\label{sec:KNIFE}
The status of previous H$_2$O maser observations with KNIFE is described in detail by \citet{ima97} and \citet{sud02}. The KNIFE observation in this paper was conducted on 1992 June 5 using the 34 m telescope of the Communication Research Laboratory (CRL) and the 45 m telescopes of the Nobeyama Radio Observatory (NRO), National Astronomical Observatory. The left-hand circular polarization signals were recorded using the K-4 VLBI backend system \citep{kiu91} at a rate of 64~Mbps in 1-bit quantization. The data correlation was made using NAOCO (New Advanced One-unit Correlator)\citep{shi94} and the Mitaka FX correlator \citep{chi91}, but we used the latter products which were provided in a FITS (Flexible Image Transport System) format. The data contained 512 spectral channels in one of 16 base band channels each with a width of 2~MHz, corresponding to a velocity spacing of 0.056~km~s$^{-1}$. 

Data reduction was performed using the Astronomical Imaging Processing System (AIPS) package in standard procedures. The visibility amplitude calibration was made using the so-called ``template spectrum method'', in which the total-power spectra taken with a reference antenna (Kashima 34 m telescope in this observation) in a specific time was used to trace temporal variations of the gains of the two telescopes. The clock offsets were calibrated using scans on 3C~273B. Fringe fitting and self-calibration were made using a bright H$_2$O maser emission at 42.61~km~s$^{-1}$ in W Hya. The CLEAN algorithm of deconvolution with uniformly weighted visibilities yielded a synthesized two-dimensional Gaussian beam to be 23$\times$5~mas at a position angle of $\approx$2\arcdeg. The 1-$\sigma$ noise level was $\approx$0.1~Jy in emission-free channels. The individual maser spots in the spectral channels were identified in fitting to a Gaussian brightness distribution. Those spots were grouped as maser features so that each group is composed of the spots located at the same physical region within a few milliarcseconds (mas) in position and $\sim$1~km~s$^{-1}$ in line-of-sight velocity. The maser spots and features identified in the present work are given in Appendix \ref{sec:spots-features}. 

\subsection{Astrometric observations with VERA}
\label{sec:VERA}

VERA has been operated mainly for trigonometry of Galactic maser sources using the dual-beam system (e.g. \cite{rei14}). Astrometric observations for W~Hya were conducted using this system by tracking J134215.3$-$290041, 1\arcdeg.62 away from W~Hya, as a position and fringe-phase reference together with W Hya. We adopted the delay-tracking center of W~Hya to be (R.A., decl.)[J2000.0]$=$(13$^{\rm h}$49$^{\rm m}$01$^{\rm s}$\hspace{-2pt}.978, $-$28$^{\circ}$22$^{\prime}$03$^{\prime\prime}$\hspace{-2pt}.81). Table \ref{tab:obs-VERA} gives the summary of the present VERA observations, including the information on the participating telescopes (Mizusawa, Iriki, Ogasawara, and Ishigakijima 20 m telescopes). The system noise temperatures of the telescopes were monitored to be typically 150--300~K, depending on weather condition and the telescope site. The maser and reference sources were observed in bandwidths of 16~MHz and 240~MHz, respectively. The received left-hand circular polarization signals were recorded at a rate of 1024~Mbps in 2-bit quantization. The data correlation was made using Mitaka FX correlator, yielding 512 spectral channels in a base band channel width of 16~MHz for the maser data, corresponding to a velocity spacing of 0.42~km~s$^{-1}$. 

The detail of the data reduction procedures are seen in, e.g., \citet{nak14}. The reference source had an observed peak intensity of 120--300~mJy depending on observation condition. In fact, due to low antenna elevations, the maser images synthesized through the phase-referencing technique were more or less defocused, causing reduction of the detected maser spots and features. For obtaining a larger number of the detected spots/features in each maser map, the maser data were self-calibrated using the bright maser spot located at the map origin. Table \ref{tab:obs-VERA} also gives comparison between the numbers of the maser features detected in the self-calibration and those in the phase-referencing procedures.  The synthesized beam sizes were typically  $\approx 1 \times 2$~mas for naturally weighted visibilities. This sharp synthesized beam was yielded by detections of compact maser features whose origin is discussed later. Even in the worst case, astrometric accuracy of 1~mas is promising (e.g. \cite{nak14}) as long as successful maser image synthesis works through the phase-referencing technique. Uncertainty of the derived absolute coordinates of the maser source may be equal to this amount because that of the position-reference quasar is smaller ($\sigma\leq$0.2~mas)\footnote{see the  VLBI global solution: http://astrogeo.org/vlbi/solutions/rfc\_2018d .}. 

\subsection{Imaging observations with KaVA}
\label{sec:KaVA}

KaVA has been scientifically operated since 2012 (e.g. \cite{yun16}) by combining the VERA and three KVN telescopes (Yonsei, Ulsan, and Tamna) and enables us to detect extended maser spots such as those in W~Hya which were spatially resolved out with long baselines ($\geq$1000~km) \citep{ima97}. W~Hya was observed with KaVA on 2013 January 8 for commissioning and science verification. The system noise temperature was typically 150--300~K in the VERA telescope, except in Ogasawara telescope with a higher temperature (700--1500~K) in the presented observation, and 70--150~K in the KVN telescopes. The received left-hand circular polarization signals were recorded at a rate of 256~Mbps in 2-bit quantization. The data correlation was made at that time using Mitaka FX correlator, yielding 1024 spectral channels in one of the two base bands each with a width of 16~MHz for the maser data, corresponding to a velocity spacing of 0.21~km~s$^{-1}$. 

The clock offsets were calibrated using scans on J133739.8$-$125724.  Fringe fitting and self-calibration were made using a bright H$_2$O maser emission at 37.56~km~s$^{-1}$ in W~Hya {\bf as reference. Because the maser emission, especially that in the reference velocity component was faded and spatially resolved out in long baselines ($>$500~km), the visibility calibration solutions were obtained from the data of the shorter baselines. Eventually, we could use the visibilities with the available calibration solutions, 
which caused data taper for the longer baselines and yielded a synthesized beam to be 27.3$\times$20.0~mas at a position angle of 15\arcdeg .1 in the final image cube}. The 1-$\sigma$ noise level was $\approx$85~mJy in emission-free channels. 


\section{Results}
\label{sec:results}

\subsection{Variation of the H$_2$O maser spectrum}
\label{sec:spectrum}
Figure \ref{fig:spectra} shows a series of total-power spectra taken in the VERA observations as example spectra of the H$_2$O masers around W~Hya. The spectral profile looks relatively simple including only one to two peaks and ttemporally stable. \citet{shi08} showed a clear time lag of the H$_2$O maser flux variation by $\sim$70~d  (a phase lag of $\Delta\phi\sim$0.2) with respect to the stellar light curve. However, the time/phase lag in the present variation of the maser spectra {\bf  is} much shorter than that found by \citet{shi08} or negligible. Figure \ref{fig:v-integ-flux} shows such time variation of the velocity-integrated total-power flux of the H$_2$O masers. \citet{shi08} pointed out the possibility that the phase lag should be longer by a light curve cycle.  Thus the periodic variability of the maser flux looks significantly modulated in W~Hya. 

\subsection{Variation of the maser spot/feature distribution}
\label{sec:distribution}
\begin{figure*}[p]
 \begin{center}
     \FigureFile(150mm,190mm){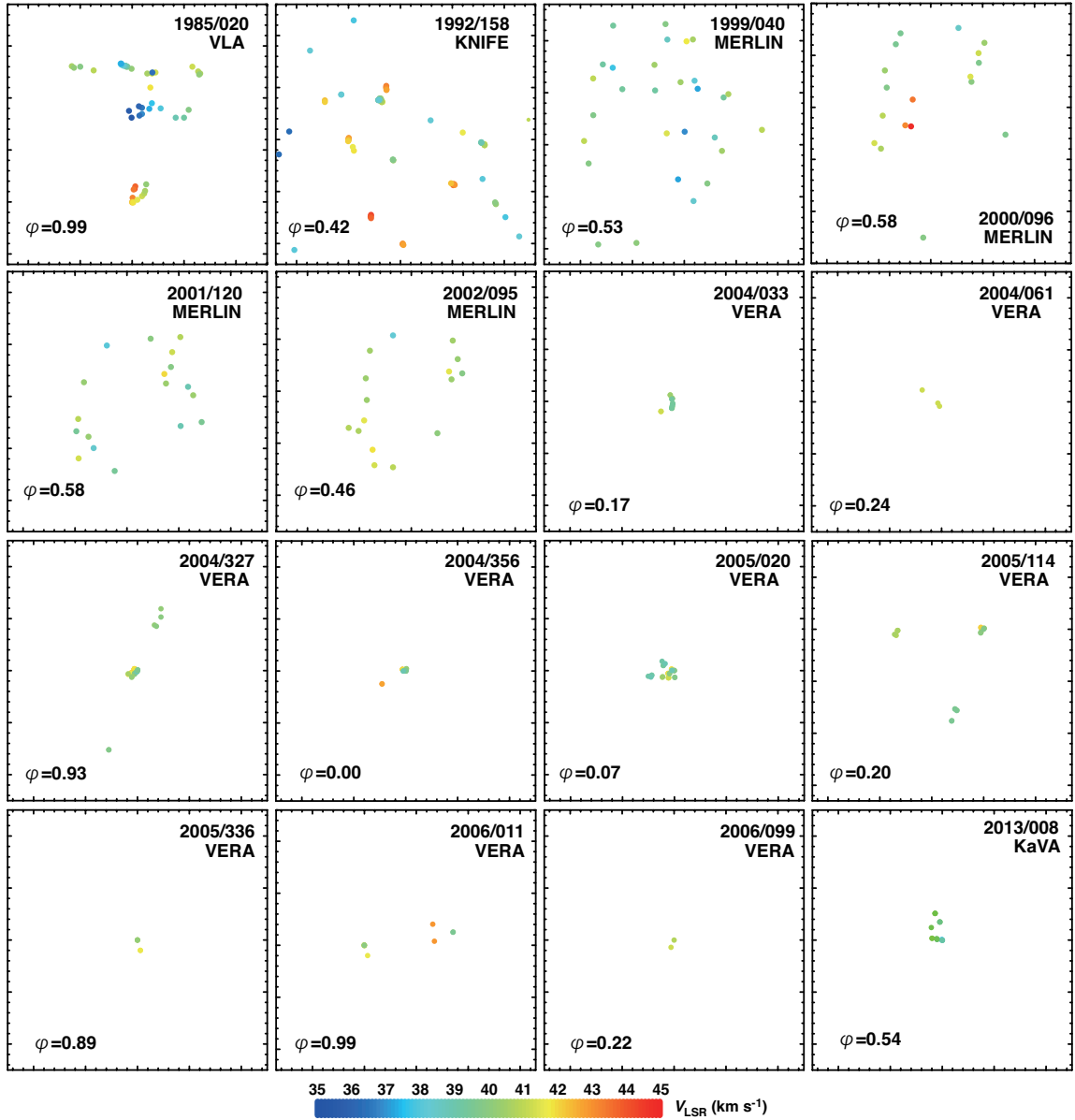}
 \end{center}
    \caption{H$_2$O maser maps of W~Hya. The interferometric observations whose results are shown in this figure 
    are listed in table \ref{tab:observations}. The displayed area is 500~mas$\times$500~mas.}
 \label{fig:maps}
\end{figure*}

The data of the identified maser spots and features are summarized in Appendix \ref{sec:spots-features}. The maser positions are obtained with respect to a reference maser spot, which was used in the fringe-fitting and self-calibration procedures and put at the map origin. There exist a larger number of maser features on the synthesized images than that of the spectral peaks mentioned above. Figure \ref{fig:maps} shows a time-series of the distributions of the H$_2$O masers revealed from all the collected data except those by \citet{rei90}. 

The oldest map of the H$_2$O masers taken with the VLA in 1985 \citep{bow93} shows the extent of maser distribution over $\sim$300~mas in the synthesized beam size of $\approx$70~mas. That observation was made around the light curve maximum. In this map, a large velocity gradient is recognized in the north--south direction but with significant deviations. The map taken with KNIFE in 1992, around the light curve minimum, shows the largest area of H$_2$O maser distribution even with only one VLBI baseline. It is difficult to find any systematic trend in the velocity distribution. Because the VLA and KNIFE observations had similar sensitivity, the difference in the apparent maser distributions should be intrinsic. 

Only the maps taken with MERLIN (yielding a typical synthesized beam size of $\approx$25~mas) in 2001 and 2002 (around the light minima) show a ring-shaped distributions of masers. The narrower velocity coverage indicates that these masers were radially expanding on the plane of sky as suggested by \citet{ric12}. The faintest maser spots in MERLIN are brighter than those in VLA and KNIFE, likely because the short interferometric baselines of MERLIN effectively could collect the flux density of the maser from extended structures.  

In the maps taken with VERA in all the sessions during 2004--2006, a smaller number of maser features than the previous VLA/KNIFE/MERLIN observations were detected. Clearly this is attributed to the characteristics of VERA that is sensitive to compact ($\leq$1~mas) maser spots. We note that the detected spots are always condensed within $\sim$100~mas with minor exceptions. 

In the maps taken with KaVA in 2013, there exist few maser features detected within 90~mas. Taking into account the available shortest baselines ($\sim$400 km from the KVN baselines) and the observation sensitivity, the relatively small extent of maser distribution should be intrinsic to the maser source. However this extent at the light curve phase of 0.54 contrasts with those found in the VLA/KNIFE/MERLIN observations conducted around the light maxima or minima.  

\begin{figure*}[p]
     \FigureFile(170mm,85mm){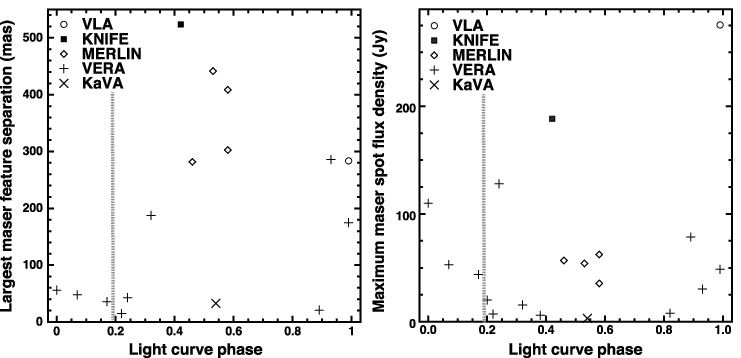}

    \caption{(a) Extent of the H$_2$O maser distribution around W~Hya against the stellar light curve phase. 
    The vertical broken line indicates the offset phase ($\Delta\phi\simeq$0.19) of the maser flux maximum with respect to 
    the optical light curve maximum, which is derived by \citet{shi08}. (b) Same as (a) but 
    the vertical scale shows the peak flux density of the maser in the synthesized image cube.}
 \label{fig:distribution-phase}
 
  \FigureFile(83mm,85mm){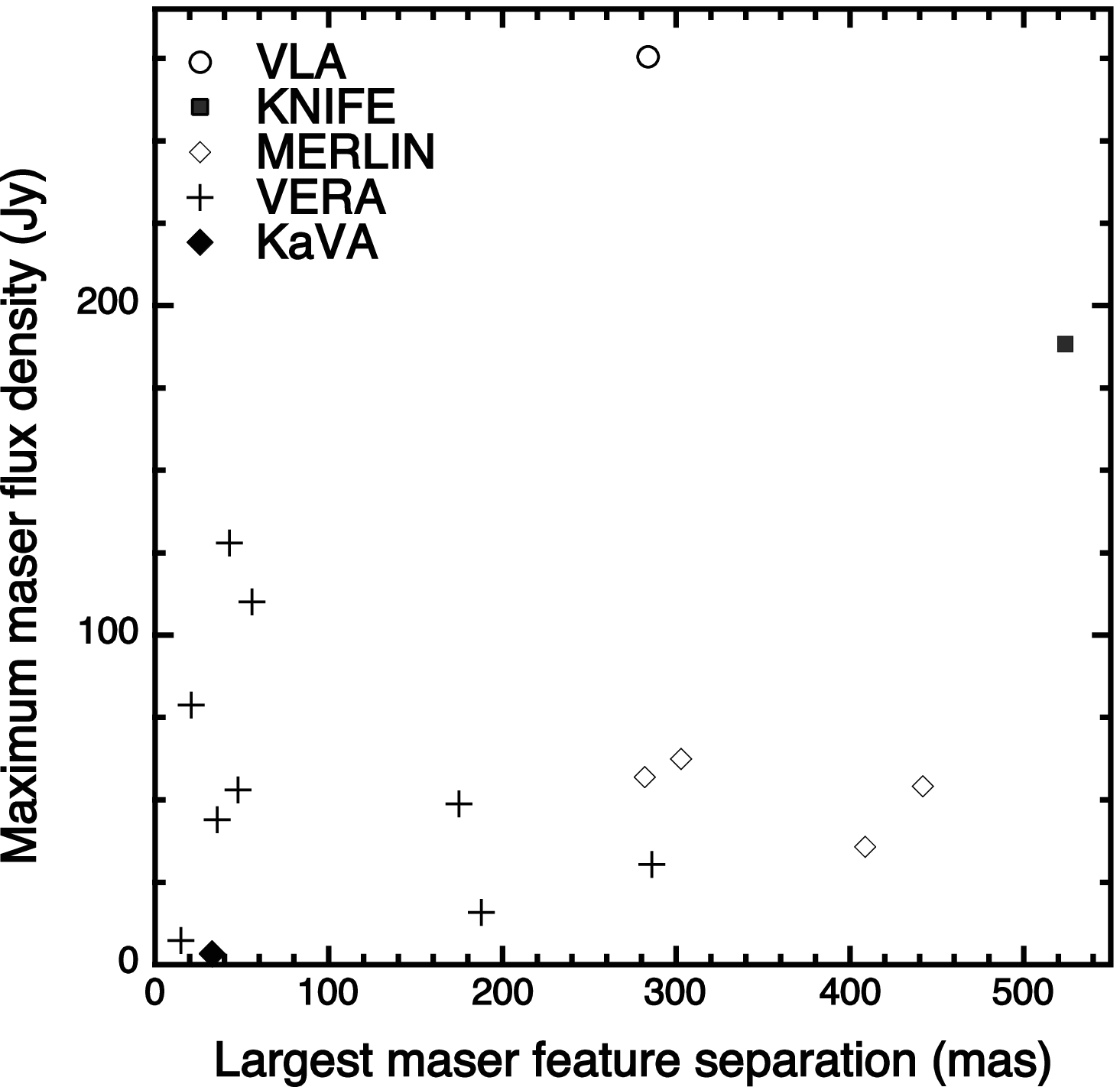}
    \caption{Extent of the H$_2$O maser distribution around W~Hya against the peak flux density of the maser in the synthesized image cube.}
 \label{fig:distribution-flux}
\end{figure*}

Figure \ref{fig:distribution-phase} shows the extent of the spot distribution and the maximum flux density of the H$_2$O masers against the stellar light curve phase. Note that, as summarized in Table \ref{tab:observations}, the sensitivities of the VLA/ MERLIN/KaVA observations and those of the VERA observations in good conditions were roughly equal. This may be attributed to the integration time much shorter with the VLA and MERLIN than that with KNIFE, VERA, and KaVA. We find that the extent of the spot distribution is not affected by the observation sensitivity or the maser spot sizes. This may support an idea that the whole extent of the maser distribution is determined primarily by compact spots although the higher sensitivity of the VLA and MERLIN to extended ($>$10~mas) maser emission should be taken into account. Thus we find that the total extent seems to have a maximum around the light curve minimum and a minimum soon after the light maximum, roughly consistent with the time lag of the variation in the total maser flux suggested by \citet{shi08}. On the other hand, it is difficult to find systematic trend of the time variation in the maximum maser spot flux density. Different from the total maser flux as shown in Figure \ref{fig:spectra} and suggested by \citet{shi08}, the individual compact maser spots visible in our VLBI observations do not seem to have correlation with the light curve. 

\begin{figure*}[p]
     \FigureFile(85mm,85mm){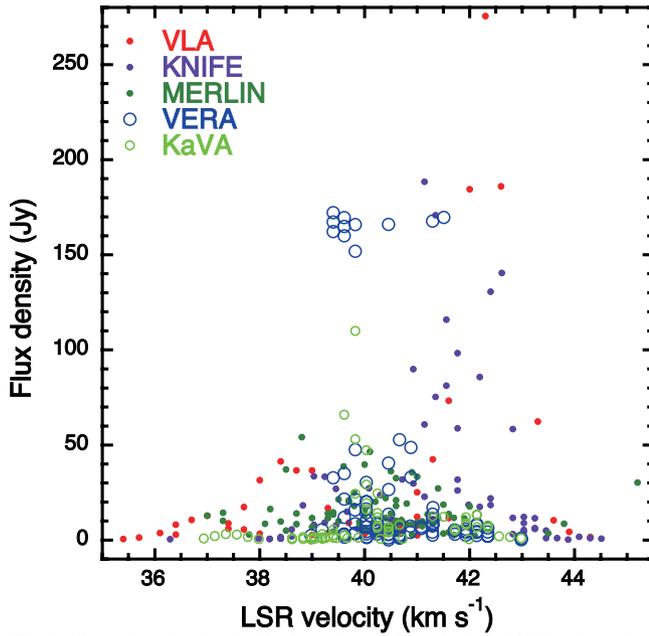}
    \caption{Maser flux density distributions along LSR velocity in W~Hya. Intensity (Jy~beam$^{-1}$) is plotted for the VLA data.}
 \label{fig:flux-vlsr}
 \end{figure*}

Figure \ref{fig:distribution-flux} shows the extent of the H$_2$O maser distribution against the maximum flux density of the H$_2$O masers. It is difficult to see correlation of the maximum maser flux density with the extent of the maser distribution. However, it is important to note that bright maser spots may contribute to extension of the apparent maser distribution. Figure \ref{fig:flux-vlsr} shows the maser flux density distribution along line-of-sight velocity with respect to the local standard of rest (LSR). Bright maser spots tend to appear in $V_{\rm LSR}\simeq$39.5--42.5~km~s$^{-1}$, slightly red-shifted with respect to the stellar systemic velocity ($V_{\rm sys}=$39.2~km~s$^{-1}$, \cite{vle17}) and big offsets from the center of the maser distribution. Supposing uniform radial expansion of a CSE hosting H$_2$O masers, the bright maser spots are expected to be located along a ring around the central star \citep{rei90}. In fact, the MERLIN maps show such a ring-shaped maser distribution (Figure \ref{fig:maps}). However, some deviations from the stellar velocity for bright maser spots as shown in Figure \ref{fig:distribution-flux} implies large deviation from such a simple case due to possible inhomogeneity of the stellar mass loss flow as discussed later.  

\subsection{Sizes and flux densities of maser spots} 
\label{sec:spot-flux}
\begin{figure*}[p]
     \FigureFile(85mm,85mm){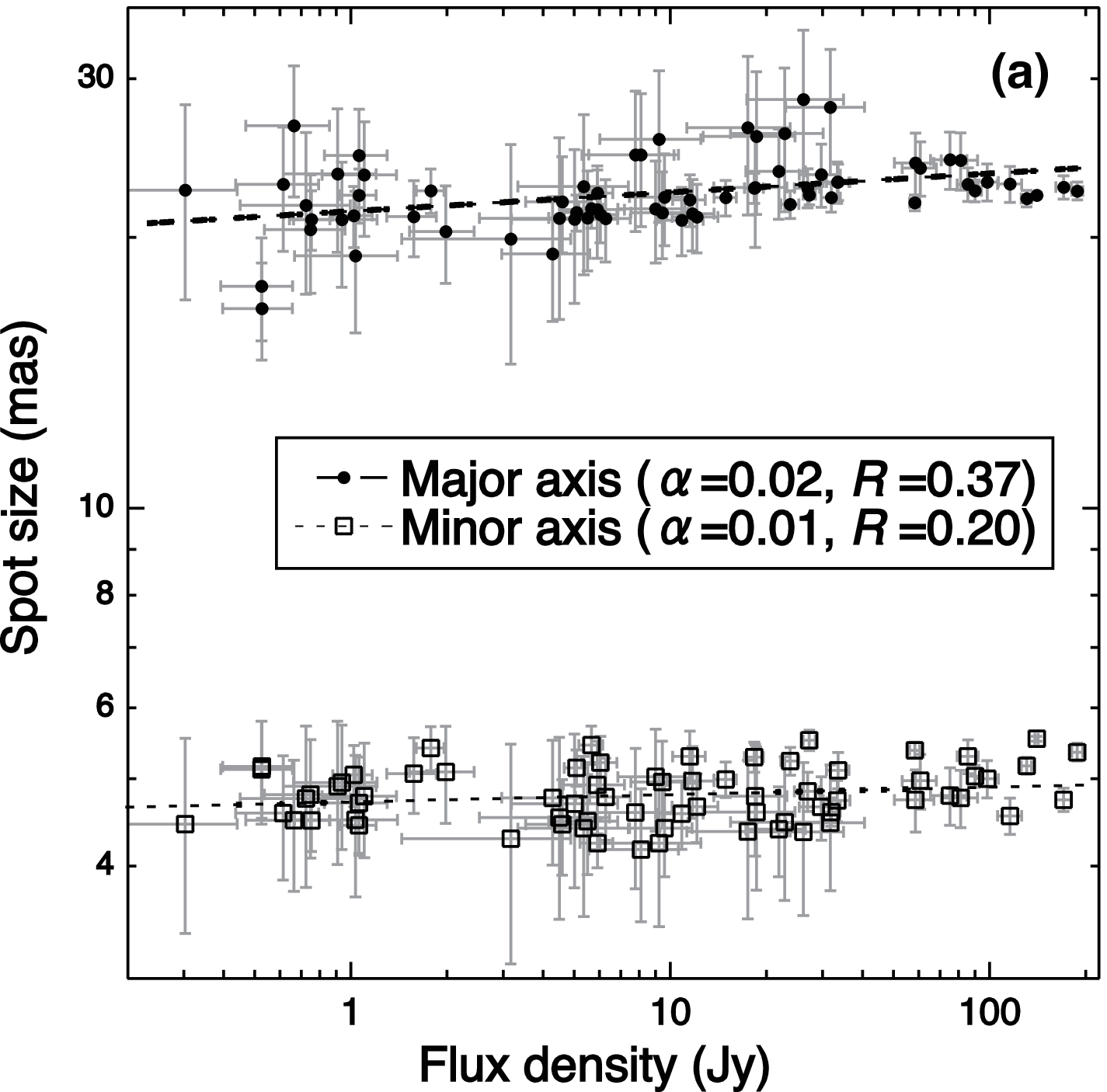}
     \FigureFile(85mm,85mm){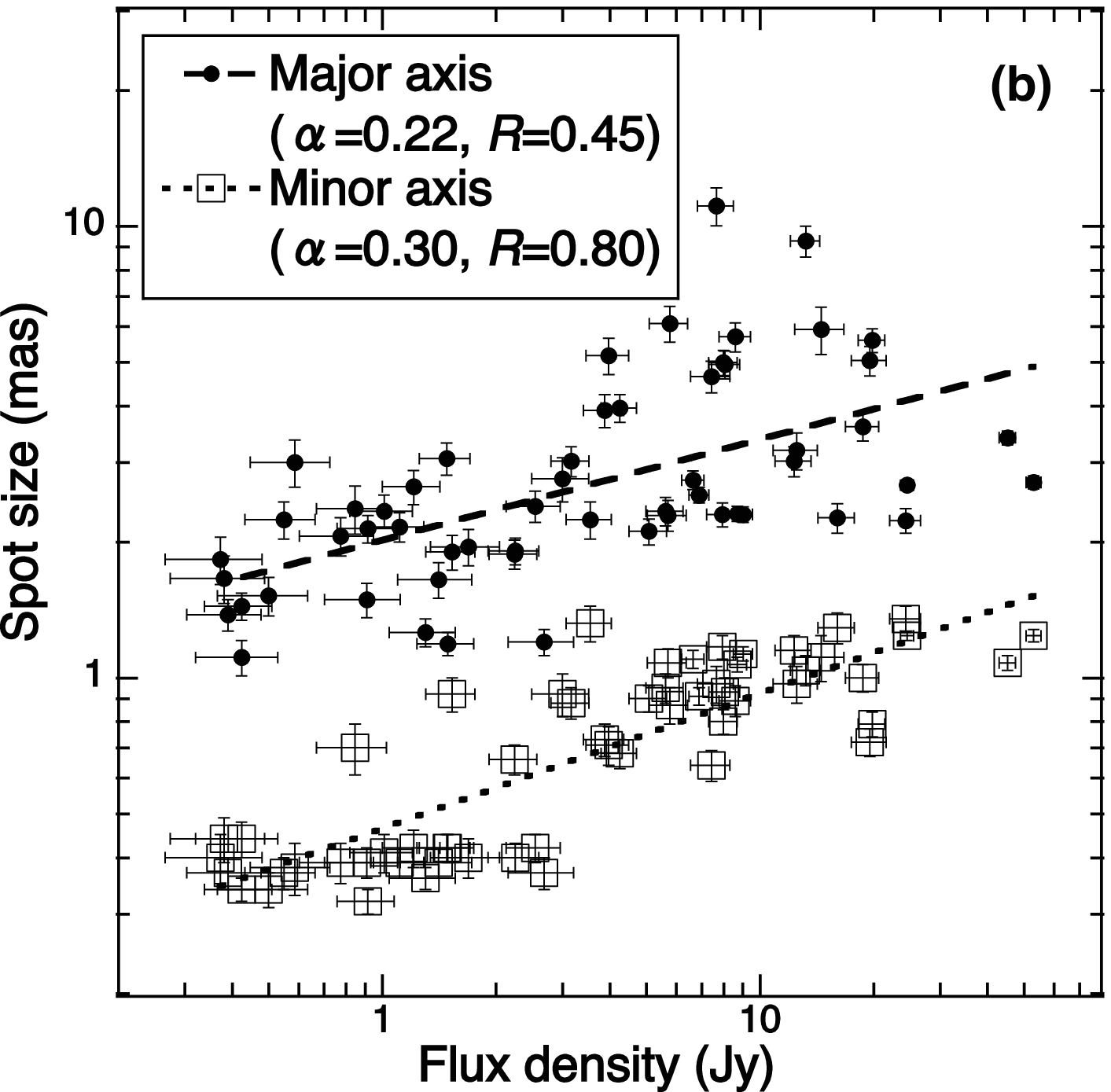}
     \vspace{3mm}
    \caption{(a) Spot size against the flux density of the H$_2$O masers around W~Hya For the KNIFE observation data. Linear fit lines ($\Theta \approx F_{\nu}^{\alpha}$) are also displayed for the major and minor axes of the maser spots that are derived from Gaussian brightness fitting. (b) Same as (a) but for the VERA observation data taken in the session on 2005 January 20.}
 \label{fig:knife-size}
 \end{figure*}

It is the most straightfoward to directly image shapes of masers spots with a sufficiently large number ($>>$10) of VLBI baselines for investigating maser beaming. Instead, \citet{ric11} proposed a method to distinguish between ``matter-bounded" and ``amplification-bounded" beaming of the individual maser features by investigating correlation between spot size and flux density. A negative or no correlation and a positive correlation are expected, respectively, for those beamings. Such an investigation may be valid with an array including short and intermediate-length baselines (say 10--500~km) which can resolve the masers into individual maser features but still recover correlated flux density from extended structures. Figure \ref{fig:knife-size} shows the result of such correlation for the W Hya H$_2$O masers using our data taken with KNIFE and VERA. The baseline length of $\sim$200~km in KNIFE is suitable for such investigation. The absence of the considered correlation in the KNIFE data suggests the matter-bounded beaming. For comparison, we can clearly see the positive correlation for the VERA data, suggesting that the maser spots detected with the VERA baselines (1000--2300~km) are amplification-bounded and bullet-shaped while the extended matter-bounded spot components are resolved out. 

\subsection{Maser spot locations with respect to the star} 
\label{sec:star-position}

\begin{figure}[p]
     \FigureFile(85mm,85mm){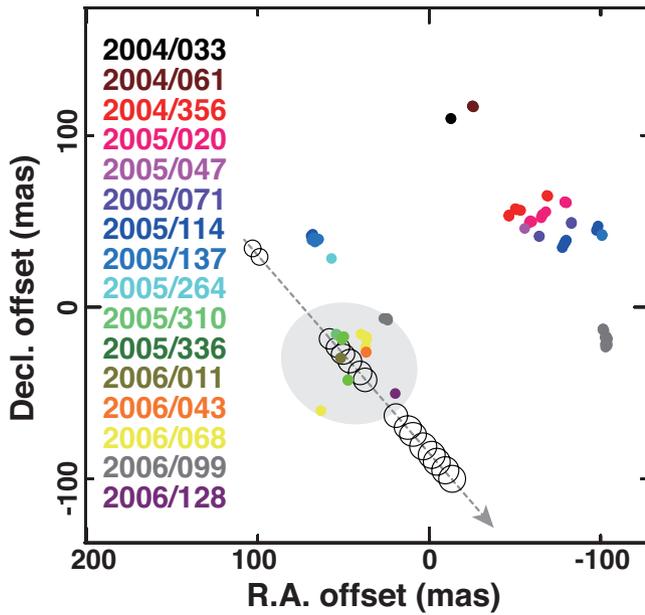}
     \caption{Time variation of the H$_2$O maser distribution around W~Hya, which is found in the VERA astrometric observations. 
     A black dashed-line arrow indicates the proper motion of the central star estimated from the {\it Gaia Data Release 2} 
     \citep{gaia18}. 
     An opened ellipse indicates the estimated position of the central star at a specific epoch of the VERA observation. 
     The size of the ellipse indicates the position uncertainty. A light-grey filled ellipse indicates the appearance of 
     the SiO maser ring  on 2005 March 12, whose size is estimated from the SiO maser distribution found by \citet{ima10}.}
 \label{fig:map-time}
\end{figure}

\begin{figure}[p]
     \FigureFile(85mm,85mm){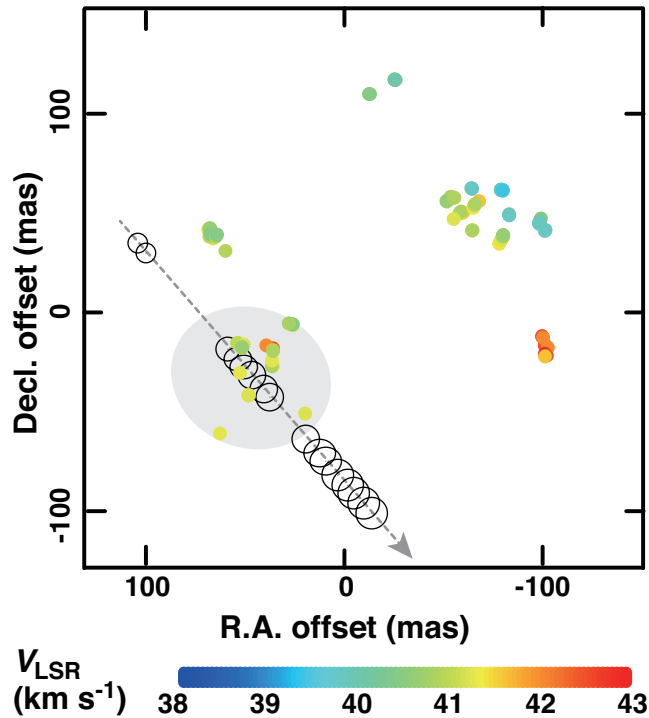}
    \caption{Same as Figure \ref{fig:map-time} but the LSR velocity distribution of the masers found at all the epochs.}
 \label{fig:map-vel}
\end{figure}

Our astrometric observations enable us to compare the maser distributions with the stellar position estimated with the stellar kinematic parameters derived by \citet{gaia18}. Using {\it SIMBAD} database, we obtained the information of the position at J2000.0 and the proper motion of the star,  (13$^{\rm h}$49$^{\rm m}$02$^{\rm s}$\hspace{-2pt}.0018, $-$28\arcdeg22\arcmin 03\arcsec 532) and $(\mu_{X}, \mu_{Y})=(-51.77\pm 1.30, -59.69\pm 1.26)$ [mas~yr$^{-1}$], respectively. Figure \ref{fig:map-time} and \ref{fig:map-vel} shows the locations of the H$_2$O masers with respect to the star.   The maser locations are biased to the northwest direction during the VERA observations. The lifetimes of the individual maser features are shorter than 100~d, making our trigonometry impossible. 



\section{Discussion}
\label{sec:discussion}

Here we attempt to describe the general properties of circumstellar H$_2$O masers that are implied from our analysis of the collection of maser spot/feature distributions around W~Hya over two decades with accurate astrometric information in some observation sessions. 

\subsection{Where do circumstellar H$_2$O maser features favor?}
\label{sec:maser-favor}
Recently W~Hya has been observed using the state-of-the-art astronomical facilities such as Very Large Telescope Interferometer (VLTI) and the Atacama Large Millimeter-and-submillimeter Array (ALMA) in excellent angular resolution ($\leq$50~mas) for thermal molecular and dust emission. All of those observations have shown that thermal emissions in the CSE is biased to north and south directions from the central star on a $\sim$50~mas scale \citep{ohn17, tak17, vle17, kho19}. The derived distribution of H$_2$O masers is also significantly biased up to $\sim$150~mas northwest of the star (Section \ref{sec:star-position}). This bias is roughly consistent with one of the three peaks of TiO emission with the continuum emission subtracted \citep{ohn17}. This may suggest that dust condensation would be enhanced behind the intensive thermal emission region then outward acceleration would be efficiently triggered  toward the outer H$_2$O maser regions, leading to maser excitation by the outward shock propagation. Theoretical works (e.g. \cite{hof03}) suggest such outward acceleration with shock wave propagation although the amplitude of the wave decreases outward. 
The suggested acceleration is probably traced by a large full velocity width of the $v=0$ $J=8\rightarrow 7$ line emission of $^{29}$SiO  molecules ($\sim$20~km~s$^{-1}$, \cite{tak17}), which are visible before their complete depletion onto dust in the H$_2$O maser region.

We speculate that shock waves may propagate to the H$_2$O maser region to enhance the maser excitation. However, one needs to caution in direct comparison of the distribution of H$_2$O maser with those of thermal molecular emissions.  In fact, the maps of the $^{29}$SiO line and the TiO emission were taken in 2015 November--December and 2016 March, respectively \citep{tak17,ohn17}. Similarly, one can find spatial correlation of the 22~GHz H$_2$O maser emission with the excited OH and 252~GHz H$_2$O maser emission \citep{kho19}, but the latter maps were taken in 2017 October. Thus our interpretation is justified if the morphologies of the TiO, excited OH, and 252 GHz H$_2$O emission regions are stable over a few decades. It is clear that the regions of optical thermal and H$_2$O maser emission do not have any spatial correlation with the direction of the stellar proper motion, therefore the maser excitation is tightly linked to the activity of the central star rather than the interaction of the CSE with ambient interstellar medium.

\subsection{What is the major agent to control the global H$_2$O maser distributions?}
\label{sec:maser-distribution-control}
There exist two major hypotheses to explain temporal variation of the global distribution of circumstellar H$_2$O masers: 1) periodic propagations of shock waves driven by stellar pulsation (e.g., \cite{ima03,shi08}) and 2) periodic variation of the physical condition, such as gas and dust temperature, in the H$_2$O maser region so as to control efficiency of maser excitation \citep{ric10,ric12}. The former better explains the phase lag observed in the variation of H$_2$O maser flux density and the short lifetimes of the H$_2$O maser features that are excited only in passage of shock waves. The latter better explains the clearer and more rapid change in the extent of maser distribution than the speed of shock propagation. The observed time lag also may be explained by a timescale of dust/gas heating by stellar radiation. 

\begin{figure*}[p]
     \FigureFile(180mm,90mm){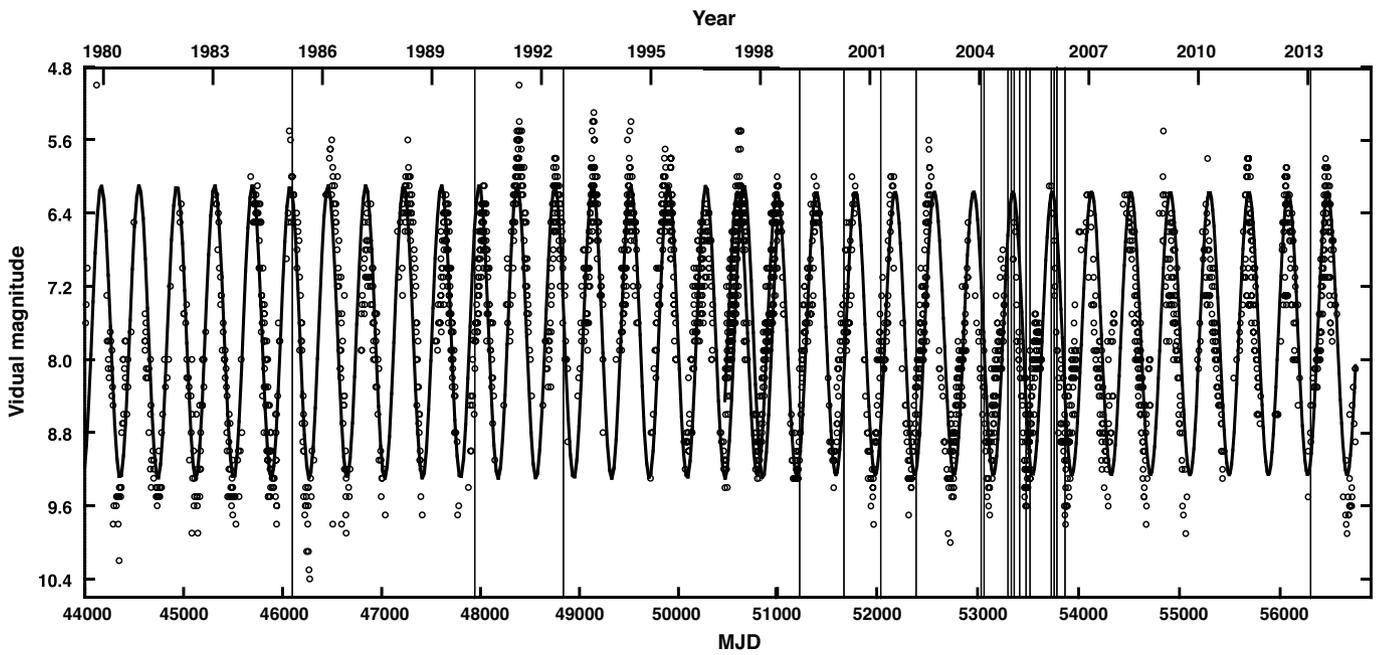}
    \caption{Visual light photometric data of W~Hya collected by AAVSO. Two solid curves show the light curves fitted to the photometric data using sin functions (see the main text). A vertical thin and solid line indicates the epoch of the interferometric observation cited in this paper. Vertical grids in dotted lines are shown in every 1000 d.}
 \label{fig:lightcurve}
\end{figure*}

In the case of the H$_2$O masers around W Hya, the former case may be more plausible for the following reasons. The first, the star is a semi-regular (SRa) variable with a relatively small amplitude of the light curve (see also Figure \ref{fig:lightcurve}), making it difficult to control the physical condition of the large volume of the CSE including the H$_2$O maser region. The second, if the effective acceleration is true as mentioned in Section \ref{sec:maser-favor}, the H$_2$O maser region may be optically obscured behind the acceleration region close to the central star and less heated.  The third, as described in Section \ref{sec:spot-flux}, the maser beaming may be matter-bounded, suggesting the maser gas clumps somehow compressed by shocks. One problem of this hypothesis, however, is that it is difficult to explain sustainable excitation of the H$_2$O masers in W~Hya that are in a class of the brightest circumstellar H$_2$O masers in spite of relatively regular temporal variation \citep{shi08}.   

\subsection{Why are the H$_2$O masers in W~Hya short-lived?}
\label{sec:maser-lifetime}
It is likely that  a maser gas clump with ``matter-bounded" beaming suggested in the case of shock propagation is geometrically less stable than that with ``amplification-bounded" beaming. It is noteworthy that H$_2$O masers in S Per (maybe those around red supergiants such as VX Sgr and VY CMa) and RT Vir are considered to be amplitude-bounded beaming \citep{ric11} and provided opportunities of successful trigonometry with their long-lived H$_2$O maser features \citep{cho08,asa10,zha17,xu18}. The VERA trigonometry for circumstellar H$_2$O maser sources has also been biased to those associated with Mira and OH/IR stars with longer pulsation periods and amplitudes, except for a small number of semi-regular variables, although they are relatively faint (e.g., \cite{nak14}).  In W~Hya, H$_2$O masers may be more randomly excited by turbulence. The proximity of W~Hya also likely makes it difficult to continuously trace the same maser spots from one epoch to another because even in the same seat of maser features apparent maser motions will be too fast and cause so-called a "Christmas tree effect" among several nearby maser features (Figure \ref{fig:map-time}
 and \ref{fig:map-vel}). 

\subsection{Forthcoming perspectives}
It is very clear that intensive (biweekly--monthly) interferometric monitoring observations are indispensable to reliably trace the performances of the individual H$_2$O maser features especially around semi-regular and Mira variable stars and reveal the true view of the physical conditions and dynamics of the CSE exciting these masers. Therefore, we expect new data yielded by long-term intensive VLBI  monitoring observations of circumstellar masers such as ongoing ESTEMA (EAVN Synthesis of sTEllar Maser Animations)\footnote
{See KaVA web page: https://radio.kasi.re.kr/kava/large\_programs.php\#sh1 .}. In such a project, co-locations with thermal line emission sources in sufficiently high angular resolution ($\sim$10~mas) will be a key for understanding the major agent of the H$_2$O maser excitation, but we may have to wait for a new decade when such co-locations will come true. 

\bigskip
We thank an anonymous referee for kindly providing comments helpful for improving this paper. We acknowledge all staff members and students who have helped for array operation and data correlation of KNIFE, VERA, and KaVA. We acknowledge the AAVSO for kindly providing the International Database of the optical photometric data. HI was supported by KAKENHI program (16H02167) supported by the Japan Society for the Promotion of Science (JSPS).  

\appendix
\section{Estimates of the light curve phases of W~Hya in the interferometric observations} 
\label{sec:light-curve}

We used the optical photometric data provided by the American Association of Variable Star Observers (AAVSO) to determine the stellar pulsation period of W Hya and the light curve phases at the epochs of interferometric observations. Because it was difficult to obtain a single pulsation period over the time span of the data for 35 years, the data were split into the sets before the modified Julian date (MJD) 51 000~d and after MJD 44 000~d. Figure \ref{fig:lightcurve} shows the photometric data points and the fitted sinusoidal curves. The obtained sin function has a period, a mean magnitude, an amplitude, and a zero-phase offset to be ($P$, $m_{\rm 0}$, $A$, $MJD_{\phi={\rm 0}}$)$=$($382.0\pm0.1$ d, 7.70$\pm$0.02~mag, 1.60$\pm$0.12, 44179$\pm$57~d) during MJD 44 000~d--51 000~d and ($P$, $m_{\rm 0}$, $A$, $MJD_{\phi={\rm 0}}$)$=$($389.8\pm0.2$ d, 7.71$\pm$0.02~mag, 1.54$\pm$0.03, 55310$\pm$50~d) during MJD of 50 000~d--57 000~d. Thus the mean stellar pulsation period is $\approx$386~d. 

\section{H$_2$O maser spots and features detected in the KNIFE, VERA, and KaVA observations}
\label{sec:spots-features}
\begin{figure*}[p]
 \begin{center}
     \FigureFile(150mm,150mm){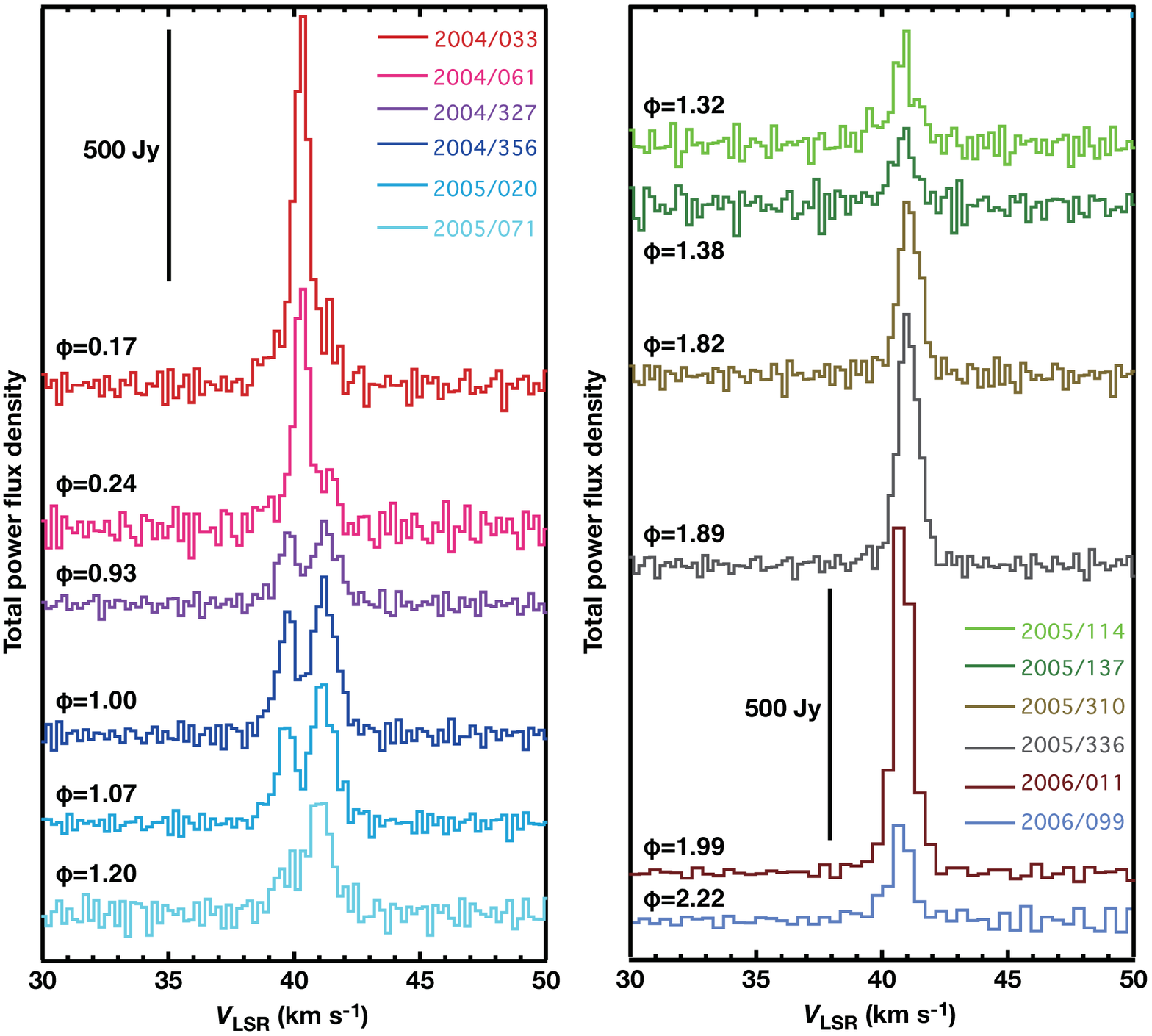}
 \end{center}
    \caption{Total power spectra of the H$_2$O maser in W~Hya, which were synthesized by 
    integrating the data from all the used telescopes of the VERA. 
    For comparison, the spectra are shifted in the vertical scale.}
    \label{fig:spectra}

     \FigureFile(83mm,85mm){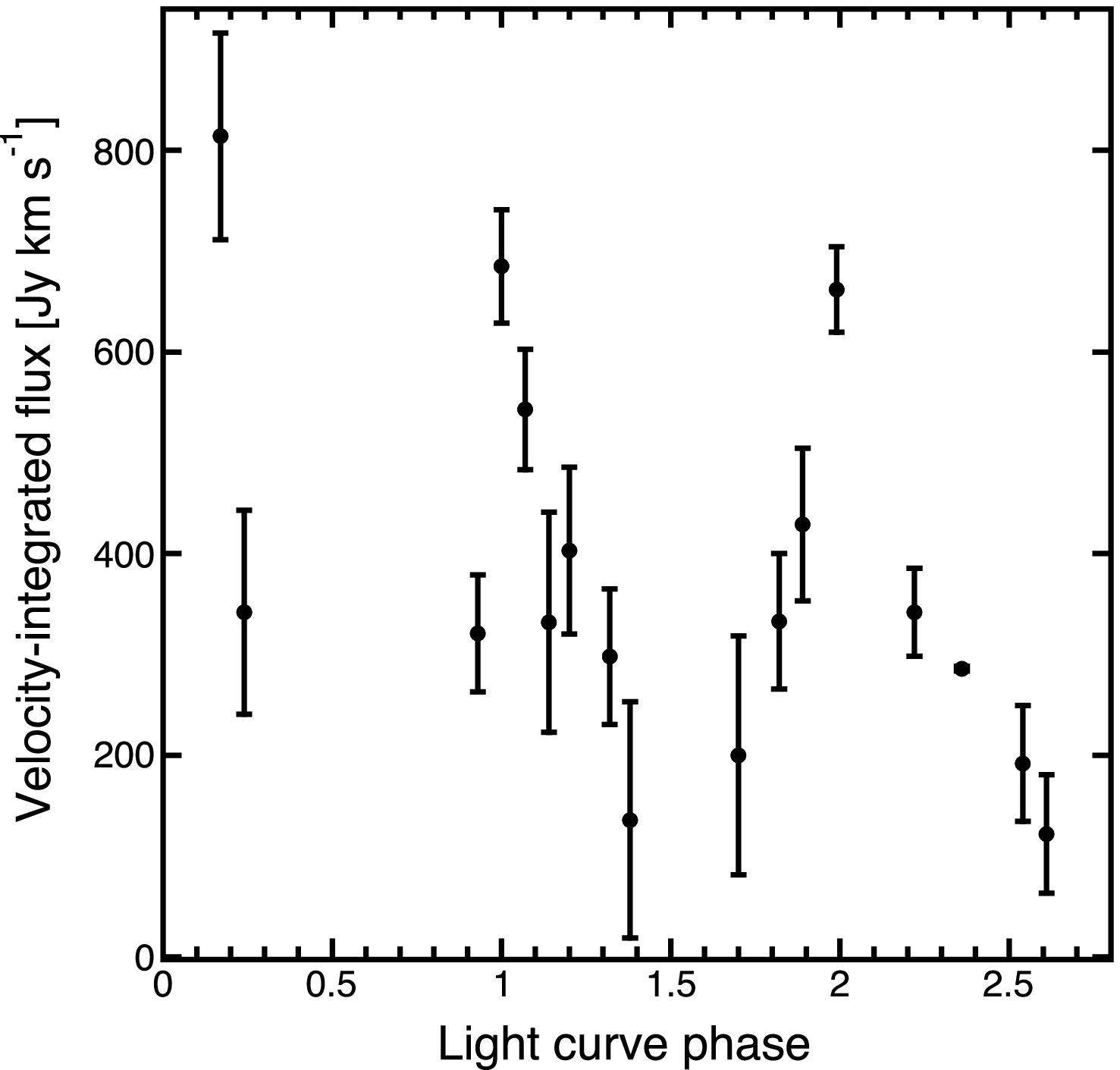}
    \caption{Time variation of the velocity-integrated total-power flux of the H$_2$O maser in W~Hya.}
 \label{fig:v-integ-flux}
\end{figure*}

Table \ref{tab:knife}, \ref{tab:vera}, and \ref{tab:KaVA} give the lists of the H$_2$O maser spots detected in the present KNIFE, VERA, and KaVA observations, respectively. For the VERA data,  a time series of the maser spectrum are also presented in Figure \ref{fig:spectra}.

\begin{table*}[t]
\caption{List of interferometric observations of H$_2$O masers in W~Hya}
\label{tab:observations}
\begin{tabular}{l@{ }r@{ }l@{ }l@{ }r@{ }r} \hline \hline
Date\footnotemark[1] & & & & & \\
(yyyy/mm/dd [DOY]) & MJD & Phase\footnotemark[2] & Facility & $S_{\rm min}$\footnotemark[3] & Ref.\footnotemark[4] \\ \hline
1985/01/20 [020] & 46086 & 0.99 & VLA & 0.6 & 2 \\
1990/02/27 [058] & 47949 & 0.87 & VLA & N/A & 3 \\
1992/06/06 [158] & 48779 & 0.42 & KNIFE & 0.4 & 1 \\
1999/02/09 [040] & 51229 & 0.53 & MERLIN & 3.0 & 4 \\
2000/04/05 [096] & 51640 & 0.58 & MERLIN & 2.6 & 4 \\
2001/04/30 [120] & 52030 & 0.58 & MERLIN & 2.3 & 4 \\
2002/04/05 [095] & 52370 & 0.46 & MERLIN & 4.2 & 4 \\
2004/02/02 [033] & 53038 & 0.17 & VERA & 1.9 & 1 \\
2004/03/01 [061] & 53066 & 0.24 & VERA & 7.4 & 1 \\
2004/11/22 [327] & 53332 & 0.93 & VERA & 2.1 & 1 \\
2004/12/21 [356] & 53361 & 0.00 & VERA & 0.3 & 1 \\
2005/01/20 [020] & 53391 & 0.07 & VERA & 0.4 & 1 \\
2005/03/11 [070] & 53441 & 0.20 & VERA & 0.6 & 1 \\
2005/04/24 [114] & 53485 & 0.32 & VERA & 0.9 & 1 \\
2005/05/17 [137] & 53508 & 0.38 & VERA & 6.1 & 1 \\
2005/11/06 [310] & 53681 & 0.82 & VERA & 6.0 & 1 \\
2005/12/02 [336] & 53707 & 0.89 & VERA & 2.2 & 1 \\
2006/01/11 [011] & 53747 & 0.99 & VERA & 0.2 & 1 \\
2006/04/09 [099] & 53835 & 0.22 & VERA & 2.7 & 1 \\
2013/01/08 [008] & 56301 & 0.54 & KaVA & 0.5 & 1 \\ \hline
\end{tabular}

\noindent
\footnotemark[1]Start date of the observation. In the results of this paper, 
the observation in which the maser map is available is listed. Three digits in a bracket indicates the day of the year of the observation date. \\
\footnotemark[2]Light curve phase at the epoch of interferometric observation. \\
\footnotemark[3]Intensity (VLA) or flux density (other arrays) of the faintest detected maser spots in units of Jy beam$^{-1}$ or Jy, which is given so as to indicate the observation sensitivity. \\
\footnotemark[4]References, 1: This paper; 2: \citet{bow93}; 3: \cite{rei07}; 4: \citet{ric12}. 

\end{table*}

\begin{table*}[t]

\caption{Status of all the VERA observations of H$_2$O masers in W~Hya}
\label{tab:obs-VERA}
\begin{tabular}{lrlrrl} \hline \hline
Date & & VERA\footnotemark[2]  & \multicolumn{2}{c}{$N_{\rm f}$\footnotemark[3]} & \\
\cline{4-5} 
(yyyy/mm/dd) & Code\footnotemark[1]  & telescopes 
& $N_{\rm f}^{\rm sc}$ &  $N_{\rm f}^{\rm pr}$ & Remark \\ \hline
2003/10/29 & r03302a & MZ, IR, IS & --- & --- & A few scans \\
2004/02/02 & r04033b & MZ, IR, OG, IS & 8 & 1 & \\
2004/03/01 & r04061a & IR, OG, IS & 1 & 1 &  \\
2004/03/24 & r04084b & MZ, IR, OG, IS & --- & --- & Problem in data \\
2004/04/24 & r04115a & MZ, IR, OG, IS & --- & --- & Problem in data \\
2004/05/25 & r04146a & MZ, IR, OG, IS & --- & --- & Problem in data \\ 
2004/11/22 & r04327b & MZ, IR, IS & 11 & --- & No astrometric trial \\
2004/12/21 & r04356c & MZ, IS, OG, IS & 6 & 4 & \\
2005/01/20 & r05020b & MZ, IS, OG, IS & 12 & 3 & \\
2005/02/16 & r05047b & MZ, OG, IS & --- & 1 & \\
2005/03/12 & r05071b & MZ, IR, OG, IS & 2 & 2 & \\
2005/04/24 & r05114b & MZ, IR, OG, IS & 8 & 4 &  \\
2005/05/17 & r05137a & MZ, IR, OG, IS & 1 & 3 & \\
2005/09/21 & r05264a & MZ, IR, OG, IS & --- & 1 & \\
2005/11/06 & r05310b & MZ, IR, OG, IS & 1 & 2 & \\
2005/12/02 & r05336b & MZ, IR, OG, IS & 2 & 2 & \\
2006/01/11 & r06011a & MZ, IR, OG, IS & 6 & 1 & \\
2006/02/12 & r06043b & MZ, IR, OG, IS & --- & 1 & \\
2006/03/09 & r06068a & MZ, IR, OG, IS & --- &  3 & \\
2006/04/09 & r06099c & MZ, IR, OG, IS & 2 & 3 & \\
2006/05/08 & r06128a & MZ, IR, OG, IS & --- & 1 & \\ \hline
\end{tabular}

\noindent
\footnotemark[1]Observation code r$yyddd$x where $yy$ indicates the last two digit numbers of the observation year 
($yy=$04 for the year 2004), $ddd$ the day of the year, and $x$(a, b, c,etc.) the order of the observation in the same day. \\
\footnotemark[2]Codes of VERA telescopes whose data were available. 
MZ: Mizusawa; IR: Iriki; OG: Ogasawar; IS: Ishigakijima. \\ 
\footnotemark[3]Number of detected maser features in the image cube created through self-calibration 
($N^{\rm sc}_{\rm f}$) and that through the phase-referencing ($N^{\rm pr}_{\rm f}$). \\
\end{table*}

\begin{table}[p]
\caption{H$_2$O maser spots detected in the KNIFE observation}
\label{tab:knife}
\scriptsize
   \begin{tabular}{rr@{ }rr@{ }rr@{ }r} \hline \hline
$V_{\rm LSR}$ & $\Delta_{\rm R.A.}$ & $\sigma_{\Delta_{\rm R.A.}}$ & $\Delta_{\rm Decl.}$ & $\sigma_{\Delta_{\rm Decl.}}$ 
& $I_{\rm peak}$ & $\sigma_{I}$ \\
 (km~s$^{-1}$) & \multicolumn{2}{c}{(mas)} & \multicolumn{2}{c}{(mas)} 
& \multicolumn{2}{c}{(Jy~beam$^{-1}$)} \\ \hline
$    44.51$&$     -43.31$&   0.15 &$    -144.99$&   0.68 &         1.00 &   0.08 \\
$    44.30$&$     -43.00$&   0.13 &$    -147.07$&   0.65 &         1.34 &   0.09 \\
$    44.09$&$     -42.63$&   0.13 &$    -148.78$&   0.54 &         1.83 &   0.10 \\
$    43.88$&$     -43.38$&   0.47 &$    -151.14$&   2.34 &         0.38 &   0.09 \\
  & & & & & & \\                                                                 
$    43.67$&$    -201.71$&   0.17 &$     -87.88$&   0.70 &         1.20 &   0.09 \\
$    43.46$&$    -202.09$&   0.10 &$     -85.81$&   0.42 &         5.81 &   0.27 \\
$    43.24$&$    -203.36$&   0.15 &$     -85.91$&   0.62 &        12.40 &   0.82 \\
$    43.03$&$    -204.45$&   0.41 &$     -87.36$&   1.95 &         6.70 &   1.35 \\
  & & & & & & \\                                                                 
$    43.46$&$     -72.75$&   0.34 &$     103.36$&   1.45 &         1.51 &   0.27 \\
$    43.24$&$     -72.45$&   0.21 &$      99.98$&   0.93 &         7.86 &   0.82 \\
$    43.03$&$     -73.04$&   0.17 &$      96.92$&   0.78 &        15.50 &   1.35 \\
$    42.82$&$     -73.29$&   0.19 &$      95.17$&   0.82 &        13.92 &   1.27 \\
  & & & & & & \\                                                                 
$    43.24$&$    -104.26$&   0.26 &$    -201.17$&   1.07 &         7.00 &   0.82 \\
$    43.03$&$    -104.59$&   0.29 &$    -200.01$&   1.18 &        10.44 &   1.35 \\
$    42.82$&$    -105.55$&   0.55 &$    -202.74$&   2.31 &         4.64 &   1.27 \\
  & & & & & & \\                                                                 
$    43.03$&$      -0.55$&   0.51 &$       3.02$&   2.03 &         5.92 &   1.35 \\
$    42.82$&$      -0.07$&   0.05 &$       1.61$&   0.19 &        62.24 &   1.27 \\
$    42.61$&$       0.03$&   0.02 &$      -0.43$&   0.06 &       142.68 &   0.95 \\
$    42.40$&$       0.26$&   0.05 &$      -1.95$&   0.20 &       143.41 &   3.03 \\
$    42.19$&$       0.71$&   0.10 &$      -3.17$&   0.42 &        88.60 &   3.80 \\
  & & & & & & \\                                                                 
$    42.40$&$      45.51$&   0.32 &$      75.31$&   1.36 &        21.31 &   3.03 \\
$    42.19$&$      45.68$&   0.41 &$      72.38$&   2.24 &        18.98 &   3.80 \\
  & & & & & & \\                                                                 
$    42.40$&$    -197.61$&   0.22 &$     -84.33$&   1.15 &        26.37 &   3.03 \\
  & & & & & & \\                                                                 
$    41.77$&$     -65.41$&   0.10 &$      72.53$&   0.47 &       107.02 &   5.10 \\
$    41.56$&$     -64.68$&   0.09 &$      72.27$&   0.45 &       139.47 &   6.44 \\
$    41.35$&$     -64.20$&   0.06 &$      72.23$&   0.29 &       198.66 &   5.93 \\
$    41.14$&$     -63.98$&   0.05 &$      72.22$&   0.22 &       195.84 &   4.45 \\
$    40.93$&$     -63.89$&   0.06 &$      72.54$&   0.27 &        99.40 &   2.76 \\
$    40.72$&$     -63.77$&   0.07 &$      73.98$&   0.34 &        39.32 &   1.44 \\
$    40.51$&$     -63.21$&   0.17 &$      76.28$&   0.77 &        14.24 &   1.24 \\
$    40.30$&$     -62.09$&   0.32 &$      78.84$&   1.29 &         5.88 &   0.93 \\
  & & & & & & \\                                                                 
$    41.77$&$    -262.22$&   0.16 &$      -9.18$&   0.82 &        64.22 &   5.10 \\
$    41.56$&$    -262.21$&   0.15 &$     -10.69$&   0.76 &        87.80 &   6.44 \\
$    41.35$&$    -262.02$&   0.15 &$     -10.61$&   0.76 &        80.78 &   5.93 \\
$    41.14$&$    -261.95$&   0.15 &$     -10.59$&   0.70 &        64.17 &   4.45 \\
$    40.93$&$    -261.71$&   0.16 &$      -9.86$&   0.81 &        34.14 &   2.76 \\
$    40.72$&$    -261.36$&   0.22 &$      -9.06$&   1.10 &        12.29 &   1.44 \\
  & & & & & & \\    
$    41.98$&$     -10.29$&   0.36 &$     -21.45$&   1.99 &        19.64 &   3.55 \\
$    41.77$&$      -8.29$&   0.31 &$     -14.27$&   1.89 &        31.95 &   5.10 \\
  & & & & & & \\                                                                 
$    41.77$&$    -220.04$&   0.36 &$      13.37$&   2.34 &        26.38 &   5.10 \\
  & & & & & & \\                                                                 
\hline
\end{tabular}
\end{table}

\begin{table}[p]
\addtocounter{table}{-1}
\caption{-- continued}
\scriptsize
   \begin{tabular}{rr@{ }rr@{ }rr@{ }r} \hline \hline
$V_{\rm LSR}$ & $\Delta_{\rm R.A.}$ & $\sigma_{\Delta_{\rm R.A.}}$ & $\Delta_{\rm Decl.}$ & $\sigma_{\Delta_{\rm Decl.}}$ 
& $I_{\rm peak}$ & $\sigma_{I}$ \\
 (km~s$^{-1}$) & \multicolumn{2}{c}{(mas)} & \multicolumn{2}{c}{(mas)} 
& \multicolumn{2}{c}{(Jy~beam$^{-1}$)} \\ \hline
$    41.14$&$    -346.55$&   0.37 &$      38.22$&   2.01 &        24.50 &   4.45 \\
  & & & & & & \\                                                                 
$    40.51$&$     -86.09$&   0.24 &$     -40.31$&   1.00 &        11.21 &   1.24 \\
$    40.30$&$     -85.99$&   0.08 &$     -39.46$&   0.33 &        26.14 &   0.93 \\
$    40.08$&$     -85.70$&   0.06 &$     -39.00$&   0.25 &        27.77 &   0.72 \\
$    39.87$&$     -85.53$&   0.11 &$     -38.52$&   0.58 &         7.81 &   0.48 \\
  & & & & & & \\                                                                 
$    40.51$&$    -283.73$&   0.39 &$    -124.37$&   1.73 &         6.39 &   1.24 \\
$    40.30$&$    -283.26$&   0.24 &$    -123.23$&   1.14 &         7.32 &   0.93 \\
$    40.08$&$    -282.20$&   0.23 &$    -120.82$&   1.13 &         5.90 &   0.72 \\ 
  & & & & & & \\                                                                 
$    39.87$&$     -59.85$&   0.16 &$      78.75$&   0.63 &         6.86 &   0.48 \\
$    39.66$&$     -57.78$&   0.10 &$      76.09$&   0.43 &        16.92 &   0.77 \\
$    39.45$&$     -57.69$&   0.10 &$      75.72$&   0.48 &        30.73 &   1.54 \\
$    39.24$&$     -57.62$&   0.11 &$      75.87$&   0.52 &        38.38 &   2.04 \\
$    39.03$&$     -57.65$&   0.10 &$      76.26$&   0.45 &        35.74 &   1.66 \\
$    38.82$&$     -57.85$&   0.09 &$      76.33$&   0.37 &        19.15 &   0.73 \\
$    38.61$&$     -58.33$&   0.12 &$      75.58$&   0.45 &         6.05 &   0.30 \\
$    38.40$&$     -59.50$&   0.22 &$      76.14$&   0.86 &         1.85 &   0.18 \\
  & & & & & & \\       
$    39.45$&$    -255.85$&   0.35 &$      -6.33$&   1.87 &         8.62 &   1.54 \\
$    39.24$&$    -255.61$&   0.35 &$      -5.66$&   2.10 &        10.62 &   2.04 \\
$    39.03$&$    -255.58$&   0.30 &$      -5.10$&   1.77 &         9.84 &   1.66 \\
  & & & & & & \\                                                                 
$    38.61$&$      13.83$&   0.29 &$      86.44$&   1.06 &         2.40 &   0.30 \\
$    38.40$&$      14.07$&   0.37 &$      86.57$&   1.40 &         1.13 &   0.18 \\
$    38.40$&$      74.03$&   0.47 &$     171.03$&   1.87 &         0.88 &   0.18 \\
  & & & & & & \\                                                                 
$    38.40$&$    -328.89$&   0.47 &$    -186.34$&   1.79 &         0.99 &   0.18 \\
  & & & & & & \\                                                                 
$    38.40$&$    -157.94$&   0.36 &$      36.74$&   1.44 &         1.23 &   0.18 \\
  & & & & & & \\                                                                 
$    38.19$&$     104.13$&   0.37 &$    -212.17$&   1.87 &         0.69 &   0.12 \\
  & & & & & & \\                                                                 
$    38.19$&$    -257.91$&   0.34 &$     -75.94$&   1.53 &         0.73 &   0.12 \\
  & & & & & & \\                                                                 
$    37.98$&$     -10.07$&   0.32 &$     229.15$&   1.30 &         0.96 &   0.14 \\
  & & & & & & \\                                                                 
$    37.98$&$    -302.05$&   0.29 &$    -149.29$&   1.30 &         1.15 &   0.14 \\
  & & & & & & \\                                                                 
$    36.29$&$     134.03$&   0.27 &$     -28.58$&   0.88 &         0.77 &   0.10 \\
  & & & & & & \\                                                                 
$    36.29$&$     114.09$&   0.29 &$      15.52$&   0.98 &         0.73 &   0.10 \\                                                            
 \hline
 \end{tabular}
\end{table}

\begin{table}[p]
\caption{H$_2$O maser spots detected in the VERA astrometric observations}
\label{tab:vera}
\scriptsize
   \begin{tabular}{rr@{ }rr@{ }rr@{ }r} \hline \hline
$V_{\rm LSR}$ & $\Delta_{\rm R.A.}$ & $\sigma_{\Delta_{\rm R.A.}}$ & $\Delta_{\rm Decl.}$ & $\sigma_{\Delta_{\rm Decl.}}$ 
& $I_{\rm peak}$ & $\sigma_{I}$ \\
 (km~s$^{-1}$) & \multicolumn{2}{c}{(mas)} & \multicolumn{2}{c}{(mas)} 
& \multicolumn{2}{c}{(Jy~beam$^{-1}$)} \\ 
\hline \multicolumn{7}{c}{2004/033} \\ \hline
$    40.67$&$     -12.60$&   0.06 &$     109.94$&   0.13 &        26.70 &   3.40 \\
$    40.46$&$     -12.77$&   0.06 &$     109.88$&   0.12 &        43.90 &   5.16 \\
$    40.24$&$     -12.95$&   0.05 &$     109.85$&   0.11 &        33.30 &   3.82 \\
\hline \multicolumn{7}{c}{2004/061} \\ \hline
$    40.87$&$     -25.16$&   0.03 &$     117.26$&   0.07 &        12.10 &   0.79 \\
$    40.66$&$     -25.21$&   0.02 &$     117.20$&   0.05 &        51.90 &   2.73 \\
$    40.45$&$     -25.34$&   0.01 &$     117.13$&   0.03 &       127.00 &   4.14 \\
$    40.24$&$     -25.49$&   0.01 &$     116.89$&   0.04 &       107.00 &   3.64 \\
$    40.03$&$     -25.79$&   0.01 &$     116.83$&   0.04 &        56.60 &   2.13 \\
$    39.82$&$     -26.14$&   0.02 &$     117.03$&   0.06 &        19.80 &   1.07 \\
\hline \multicolumn{7}{c}{2004/356} \\ \hline
$    41.93$&$     -50.64$&   0.04 &$      56.88$&   0.07 &         5.26 &   0.41 \\
$    41.72$&$     -50.89$&   0.03 &$      56.99$&   0.06 &         8.18 &   0.54 \\
$    41.51$&$     -50.77$&   0.03 &$      56.98$&   0.06 &        10.00 &   0.70 \\
$    41.30$&$     -50.42$&   0.03 &$      57.15$&   0.06 &        12.60 &   0.91 \\
$    41.08$&$     -50.32$&   0.03 &$      57.31$&   0.07 &        12.80 &   0.95 \\
$    40.88$&$     -50.28$&   0.04 &$      57.37$&   0.08 &         7.01 &   0.60 \\
  & & & & & & \\                                                                 
$    41.30$&$     -46.60$&   0.03 &$      53.11$&   0.06 &        12.60 &   0.91 \\
$    41.08$&$     -46.46$&   0.03 &$      53.33$&   0.05 &        15.80 &   0.95 \\
$    40.88$&$     -46.49$&   0.02 &$      53.52$&   0.04 &        12.30 &   0.60 \\
$    40.66$&$     -46.56$&   0.02 &$      53.69$&   0.04 &         5.52 &   0.26 \\
  & & & & & & \\                                                                 
$    41.08$&$     -53.24$&   0.03 &$      56.45$&   0.06 &        15.00 &   0.95 \\
$    40.88$&$     -53.13$&   0.03 &$      56.36$&   0.06 &         9.25 &   0.60 \\
  & & & & & & \\                                                                 
$    40.24$&$     -68.63$&   0.03 &$      65.09$&   0.06 &         8.99 &   0.60 \\
$    40.03$&$     -68.94$&   0.02 &$      64.87$&   0.04 &        26.90 &   1.30 \\
$    39.82$&$     -69.08$&   0.02 &$      64.76$&   0.04 &        34.70 &   1.68 \\
\hline \multicolumn{7}{c}{2005/020} \\ \hline
$    41.72$&$     -59.87$&   0.04 &$      49.88$&   0.08 &         4.10 &   0.38 \\
$    41.51$&$     -58.31$&   0.02 &$      49.27$&   0.04 &        11.20 &   0.54 \\
$    41.30$&$     -58.55$&   0.02 &$      49.61$&   0.03 &        22.50 &   0.83 \\
$    41.09$&$     -58.83$&   0.02 &$      49.94$&   0.04 &        21.70 &   0.90 \\
$    40.88$&$     -59.04$&   0.03 &$      50.31$&   0.06 &         9.61 &   0.63 \\
  & & & & & & \\                                                                 
$    41.72$&$     -68.05$&   0.04 &$      55.63$&   0.07 &         4.56 &   0.38 \\
$    41.51$&$     -65.35$&   0.03 &$      52.08$&   0.06 &         7.40 &   0.54 \\
$    41.30$&$     -65.73$&   0.04 &$      53.37$&   0.07 &        10.50 &   0.83 \\
$    41.09$&$     -65.92$&   0.03 &$      53.73$&   0.06 &        14.00 &   0.90 \\
$    40.88$&$     -66.08$&   0.03 &$      54.01$&   0.05 &        11.30 &   0.63 \\
  & & & & & & \\                                                                 
$    40.46$&$     -79.00$&   0.02 &$      61.41$&   0.04 &         4.57 &   0.23 \\
$    40.24$&$     -79.05$&   0.02 &$      61.29$&   0.04 &        12.90 &   0.55 \\
$    40.03$&$     -79.17$&   0.02 &$      61.17$&   0.04 &        25.80 &   1.07 \\
$    39.82$&$     -79.52$&   0.02 &$      61.10$&   0.03 &        30.70 &   1.15 \\
$    39.61$&$     -79.87$&   0.01 &$      61.05$&   0.03 &        22.30 &   0.74 \\
$    39.40$&$     -79.98$&   0.02 &$      60.99$&   0.03 &         8.64 &   0.34 \\
$    39.19$&$     -80.01$&   0.03 &$      61.00$&   0.05 &         2.79 &   0.16 \\
\hline
\end{tabular}
\end{table}

\begin{table}[p]
\addtocounter{table}{-1}
\caption{-- continued}
\scriptsize
   \begin{tabular}{rr@{ }rr@{ }rr@{ }r} \hline \hline
$V_{\rm LSR}$ & $\Delta_{\rm R.A.}$ & $\sigma_{\Delta_{\rm R.A.}}$ & $\Delta_{\rm Decl.}$ & $\sigma_{\Delta_{\rm Decl.}}$ 
& $I_{\rm peak}$ & $\sigma_{I}$ \\
 (km~s$^{-1}$) & \multicolumn{2}{c}{(mas)} & \multicolumn{2}{c}{(mas)} 
& \multicolumn{2}{c}{(Jy~beam$^{-1}$)} \\ \hline
\hline \multicolumn{7}{c}{2005/047} \\ \hline
$    41.30$&$     -55.61$&   0.05 &$      45.88$&   0.09 &         6.02 &   0.73 \\
$    41.09$&$     -55.64$&   0.06 &$      45.99$&   0.10 &         5.26 &   0.74 \\
\hline \multicolumn{7}{c}{2005/071} \\ \hline
$    41.29$&$     -64.20$&   0.05 &$      41.17$&   0.14 &         9.74 &   1.29 \\
$    41.08$&$     -64.23$&   0.06 &$      41.40$&   0.14 &        11.70 &   1.65 \\
$    40.87$&$     -64.21$&   0.06 &$      41.44$&   0.16 &         8.21 &   1.28 \\
  & & & & & & \\                                                                 
$    40.24$&$     -82.91$&   0.04 &$      49.52$&   0.11 &         7.80 &   0.85 \\
$    40.03$&$     -82.92$&   0.04 &$      49.14$&   0.10 &        14.70 &   1.48 \\
$    39.82$&$     -82.92$&   0.05 &$      48.85$&   0.12 &        13.10 &   1.48 \\
$    39.61$&$     -82.96$&   0.05 &$      48.75$&   0.13 &         6.97 &   0.87 \\
\hline \multicolumn{7}{c}{2005/114} \\ \hline
$    41.50$&$     -77.52$&   0.12 &$      34.87$&   0.22 &         3.47 &   0.27 \\
$    41.29$&$     -78.61$&   0.11 &$      36.75$&   0.20 &         6.69 &   0.47 \\
$    41.08$&$     -79.32$&   0.11 &$      37.99$&   0.21 &         9.28 &   0.67 \\
$    40.87$&$     -79.59$&   0.11 &$      38.52$&   0.22 &         7.49 &   0.57 \\
$    40.66$&$     -79.85$&   0.14 &$      38.96$&   0.26 &         3.65 &   0.34 \\
  & & & & & & \\                                                                 
$    41.29$&$      68.88$&   0.25 &$      41.48$&   0.48 &         2.84 &   0.47 \\
$    41.08$&$      68.38$&   0.22 &$      42.06$&   0.41 &         4.68 &   0.67 \\
$    40.87$&$      68.13$&   0.17 &$      42.17$&   0.32 &         5.00 &   0.57 \\
$    40.66$&$      67.88$&   0.14 &$      42.42$&   0.26 &         3.61 &   0.34 \\
  & & & & & & \\                                                                 
$    41.50$&$     -77.52$&   0.12 &$      34.87$&   0.22 &         3.47 &   0.27 \\
$    41.29$&$     -78.61$&   0.11 &$      36.75$&   0.20 &         6.69 &   0.47 \\
$    41.08$&$     -79.32$&   0.11 &$      37.99$&   0.21 &         9.28 &   0.67 \\
$    40.87$&$     -79.59$&   0.11 &$      38.52$&   0.22 &         7.49 &   0.57 \\
$    40.66$&$     -79.85$&   0.14 &$      38.96$&   0.26 &         3.65 &   0.34 \\
  & & & & & & \\                                                                 
$    40.45$&$     -98.57$&   0.24 &$      47.22$&   0.46 &         1.43 &   0.23 \\
$    40.24$&$     -98.12$&   0.11 &$      46.13$&   0.21 &         4.35 &   0.33 \\
$    40.03$&$     -97.74$&   0.09 &$      45.30$&   0.17 &         8.06 &   0.48 \\
$    39.82$&$     -97.49$&   0.09 &$      44.74$&   0.17 &         8.01 &   0.46 \\
$    39.61$&$     -97.20$&   0.10 &$      44.11$&   0.19 &         4.99 &   0.33 \\
\hline \multicolumn{7}{c}{2005/137} \\ \hline
$    41.29$&$      66.74$&   0.02 &$      38.04$&   0.03 &         6.12 &   0.22 \\
$    41.08$&$      68.34$&   0.04 &$      39.09$&   0.07 &         3.33 &   0.28 \\
$    40.87$&$      68.20$&   0.03 &$      39.62$&   0.05 &         4.58 &   0.28 \\
$    40.66$&$      67.95$&   0.03 &$      40.12$&   0.06 &         3.31 &   0.23 \\
$    40.45$&$      67.61$&   0.04 &$      40.43$&   0.08 &         1.60 &   0.16 \\
  & & & & & & \\                                                                 
$    41.08$&$      65.37$&   0.03 &$      38.98$&   0.06 &         3.75 &   0.28 \\
$    40.87$&$      65.06$&   0.03 &$      39.33$&   0.06 &         3.56 &   0.28 \\
$    40.66$&$      64.73$&   0.04 &$      39.65$&   0.08 &         2.44 &   0.23 \\
$    40.45$&$      64.54$&   0.09 &$      39.65$&   0.16 &         0.80 &   0.16 \\
  & & & & & & \\                                                                 
$    40.24$&$    -100.74$&   0.05 &$      42.39$&   0.09 &         1.38 &   0.15 \\
$    40.03$&$    -100.78$&   0.03 &$      42.20$&   0.05 &         2.81 &   0.17 \\
$    39.81$&$    -100.71$&   0.03 &$      41.82$&   0.06 &         2.25 &   0.18 \\
$    39.60$&$    -100.73$&   0.06 &$      41.45$&   0.10 &         1.17 &   0.15 \\
\hline
\end{tabular}
\end{table}

\begin{table}[p]
\addtocounter{table}{-1}
\caption{-- continued}
\scriptsize
   \begin{tabular}{rr@{ }rr@{ }rr@{ }r} \hline \hline
$V_{\rm LSR}$ & $\Delta_{\rm R.A.}$ & $\sigma_{\Delta_{\rm R.A.}}$ & $\Delta_{\rm Decl.}$ & $\sigma_{\Delta_{\rm Decl.}}$ 
& $I_{\rm peak}$ & $\sigma_{I}$ \\
 (km~s$^{-1}$) & \multicolumn{2}{c}{(mas)} & \multicolumn{2}{c}{(mas)} 
& \multicolumn{2}{c}{(Jy~beam$^{-1}$)} \\ \hline
\hline \multicolumn{7}{c}{2005/264} \\ \hline
$    40.97$&$      56.97$&   0.10 &$      28.45$&   0.17 &         1.32 &   0.28 \\
\hline \multicolumn{7}{c}{2005/310} \\ \hline
$    41.50$&$      53.82$&   0.08 &$     -15.91$&   0.15 &         1.59 &   0.24 \\
$    41.29$&$      53.98$&   0.07 &$     -15.88$&   0.15 &         1.86 &   0.27 \\
$    41.08$&$      54.29$&   0.08 &$     -15.77$&   0.16 &         1.60 &   0.26 \\
  & & & & & & \\                                                                 
$    40.87$&$      46.92$&   0.08 &$      16.16$&   0.15 &         1.56 &   0.24 \\
\hline \multicolumn{7}{c}{2005/336} \\ \hline
$    41.72$&$      47.49$&   0.03 &$     -42.51$&   0.06 &         5.57 &   0.37 \\
$    41.51$&$      47.49$&   0.03 &$     -42.42$&   0.06 &         7.15 &   0.50 \\
$    41.29$&$      47.47$&   0.05 &$     -42.26$&   0.10 &         4.49 &   0.50 \\
  & & & & & & \\                                                                 
$    41.29$&$      49.78$&   0.12 &$     -17.04$&   0.24 &         1.90 &   0.50 \\
$    41.08$&$      50.97$&   0.04 &$     -18.67$&   0.08 &         6.33 &   0.56 \\
$    40.87$&$      50.83$&   0.03 &$     -18.45$&   0.05 &        13.70 &   0.78 \\
$    40.66$&$      50.76$&   0.02 &$     -18.34$&   0.05 &        12.80 &   0.69 \\
$    40.45$&$      50.66$&   0.02 &$     -18.20$&   0.05 &         6.24 &   0.33 \\
\hline \multicolumn{7}{c}{2006/011} \\ \hline
$    41.30$&$      51.80$&   0.02 &$     -29.68$&   0.03 &        17.80 &   0.68 \\
$    40.88$&$      51.73$&   0.02 &$     -29.60$&   0.03 &        50.40 &   1.96 \\
\hline \multicolumn{7}{c}{2006/043} \\ \hline
$    41.30$&$      36.77$&   0.05 &$     -26.13$&   0.08 &         2.56 &   0.30 \\
$    40.88$&$      36.88$&   0.05 &$     -26.20$&   0.08 &         5.01 &   0.57 \\
$    40.46$&$      36.87$&   0.05 &$     -26.22$&   0.08 &         2.13 &   0.23 \\
\hline \multicolumn{7}{c}{2006/068} \\ \hline
$    42.56$&$      36.71$&   0.78 &$     -17.25$&   1.41 &         0.21 &   0.09 \\
$    42.14$&$      39.83$&   0.63 &$     -15.75$&   1.13 &         0.25 &   0.09 \\
$    41.72$&$      36.56$&   0.47 &$     -18.08$&   0.84 &         0.37 &   0.10 \\
  & & & & & & \\                                                                 
$    41.30$&$      63.25$&   0.26 &$     -60.22$&   0.47 &         1.02 &   0.14 \\
  & & & & & & \\                                                                 
$    41.30$&$      37.16$&   0.12 &$     -23.38$&   0.22 &         2.20 &   0.14 \\
  & & & & & & \\                                                                 
$    40.88$&$      36.62$&   0.16 &$     -18.89$&   0.28 &         3.29 &   0.28 \\
$    40.46$&$      36.10$&   0.16 &$     -17.99$&   0.29 &         2.17 &   0.19 \\
\hline \multicolumn{7}{c}{2006/099} \\ \hline
$    43.40$&$    -103.09$&   0.07 &$     -17.03$&   0.11 &         0.54 &   0.09 \\
$    42.98$&$    -102.40$&   0.04 &$     -17.32$&   0.07 &         1.00 &   0.09 \\
$    42.56$&$    -102.62$&   0.03 &$     -17.87$&   0.05 &         1.38 &   0.11 \\
$    42.14$&$    -103.85$&   0.04 &$     -18.43$&   0.06 &         1.04 &   0.09 \\
  & & & & & & \\                                                                 
$    43.40$&$    -101.25$&   0.09 &$     -12.59$&   0.14 &         0.44 &   0.09 \\
$    42.98$&$    -101.05$&   0.05 &$     -12.78$&   0.07 &         0.91 &   0.09 \\
$    42.56$&$    -101.17$&   0.05 &$     -13.18$&   0.08 &         0.94 &   0.11 \\
$    42.14$&$    -101.65$&   0.07 &$     -13.60$&   0.11 &         0.60 &   0.09 \\
  & & & & & & \\                                                                 
 \hline
\end{tabular}
\end{table}

\begin{table}[p]
\addtocounter{table}{-1}
\caption{-- continued}
\scriptsize
   \begin{tabular}{rr@{ }rr@{ }rr@{ }r} \hline \hline
$V_{\rm LSR}$ & $\Delta_{\rm R.A.}$ & $\sigma_{\Delta_{\rm R.A.}}$ & $\Delta_{\rm Decl.}$ & $\sigma_{\Delta_{\rm Decl.}}$ 
& $I_{\rm peak}$ & $\sigma_{I}$ \\
 (km~s$^{-1}$) & \multicolumn{2}{c}{(mas)} & \multicolumn{2}{c}{(mas)} 
& \multicolumn{2}{c}{(Jy~beam$^{-1}$)} \\ \hline
$    43.40$&$    -103.66$&   0.07 &$     -21.41$&   0.12 &         0.54 &   0.09 \\
$    42.98$&$    -102.47$&   0.04 &$     -21.63$&   0.07 &         0.98 &   0.09 \\
$    42.56$&$    -103.42$&   0.04 &$     -22.34$&   0.06 &         1.33 &   0.11 \\
$    41.72$&$    -102.53$&   0.04 &$     -23.21$&   0.07 &         0.98 &   0.09 \\
  & & & & & & \\                                                         
$    41.29$&$      24.45$&   0.07 &$      -7.43$&   0.11 &         0.82 &   0.12 \\
$    40.87$&$      24.71$&   0.03 &$      -6.96$&   0.05 &         2.03 &   0.15 \\
$    40.45$&$      24.80$&   0.06 &$      -6.60$&   0.09 &         0.86 &   0.11 \\
  & & & & & & \\                                                                 
$    40.87$&$      26.75$&   0.04 &$      -6.50$&   0.06 &         1.79 &   0.15 \\
$    40.45$&$      26.77$&   0.03 &$      -6.32$&   0.05 &         1.68 &   0.11 \\
\hline \multicolumn{7}{c}{2006/128} \\ \hline
$    41.29$&$      19.89$&   0.04 &$     -50.29$&   0.08 &         1.31 &   0.11 \\
$    40.87$&$      19.82$&   0.05 &$     -50.72$&   0.09 &         1.13 &   0.11 \\                                              
 \hline
 \end{tabular}
\end{table}

\begin{table}[p]
\caption{H$_2$O maser spots detected in the KaVA observation}
\label{tab:KaVA}
\scriptsize
   \begin{tabular}{rr@{ }rr@{ }rr@{ }r} \hline \hline
$V_{\rm LSR}$ & $\Delta_{\rm R.A.}$ & $\sigma_{\Delta_{\rm R.A.}}$ & $\Delta_{\rm Decl.}$ & $\sigma_{\Delta_{\rm Decl.}}$ 
& $I_{\rm peak}$ & $\sigma_{I}$ \\
 (km~s$^{-1}$) & \multicolumn{2}{c}{(mas)} & \multicolumn{2}{c}{(mas)} 
& \multicolumn{2}{c}{(Jy~beam$^{-1}$)} \\ \hline
$    36.93$&$       0.75$&   0.19 &$       1.12$&   0.22 &         0.82 &   0.06 \\
$    37.14$&$       0.43$&   0.15 &$       0.63$&   0.19 &         1.82 &   0.12 \\
$    37.35$&$       0.15$&   0.09 &$       0.11$&   0.11 &         2.79 &   0.10 \\
$    37.56$&$       0.03$&   0.08 &$       0.02$&   0.10 &         2.43 &   0.08 \\
$    37.78$&$      -0.18$&   0.11 &$      -0.38$&   0.13 &         1.31 &   0.06 \\
$    37.99$&$      -0.55$&   0.29 &$      -0.42$&   0.30 &         0.42 &   0.04 \\
  & & & & & & \\                                                                 
$    38.62$&$       9.90$&   0.25 &$       0.97$&   0.47 &         0.67 &   0.07 \\
$    38.83$&$      10.01$&   0.24 &$       1.41$&   0.53 &         0.93 &   0.10 \\
$    39.04$&$      10.36$&   0.28 &$       2.08$&   0.55 &         0.98 &   0.12 \\
$    39.25$&$      10.65$&   0.33 &$       2.74$&   0.70 &         0.80 &   0.12 \\
  & & & & & & \\                                                                 
$    39.04$&$       4.46$&   0.35 &$      34.83$&   0.46 &         0.82 &   0.17 \\
$    38.83$&$       4.62$&   0.39 &$      34.74$&   0.49 &         0.80 &   0.17 \\
$    38.62$&$       4.83$&   0.43 &$      34.60$&   0.61 &         0.58 &   0.13 \\
  & & & & & & \\                                                                 
$    39.46$&$      21.14$&   0.37 &$      24.19$&   0.33 &         0.68 &   0.08 \\
  & & & & & & \\                                                                 
$    39.25$&$      13.22$&   0.89 &$      51.57$&   0.70 &         0.55 &   0.11 \\
$    39.67$&$      13.85$&   0.24 &$      51.30$&   0.24 &         0.74 &   0.06 \\
$    39.88$&$      13.77$&   0.41 &$      50.92$&   0.62 &         0.74 &   0.12 \\
  & & & & & & \\                                                                 
$    40.30$&$      19.55$&   0.44 &$       3.43$&   0.85 &         0.47 &   0.09 \\
$    40.72$&$      19.56$&   0.44 &$       3.48$&   0.69 &         0.38 &   0.06 \\                        
 \hline
 \end{tabular}
\end{table}

\clearpage

\begin{thebibliography}{99}
\bibitem[Asaki et~al.(2010)]{asa10} Asaki,~Y.,  et~al.\ 2010, \apj, 721, 267
\bibitem[Bowers et~al.(1993)]{bow93} Bowers,~P.~F., Claussen,~M.~J., \& Johnston, K.~J.\ 1993, \aj, 105, 284
\bibitem[Bowers et~al.(1994)]{bow94} Bowers,~P.~F., \& Johnston, K.~J.\ 1994, \apjs, 92, 189
\bibitem[Chikada et~al.(1991)]{chi91} Chikada, Y.,  et~al.\ 1991, in Frontiers of VLBI, ed. H. Hirabayashi, M. Inoue, \& H. Kobayashi (Tokyo: Universal Academy Press), p~79
\bibitem[Choi et~al.(2008)]{cho08} Choi,~Y.-K.,  et~al.\, \pasj, 60, 1007
\bibitem[Gaia Collaboration(2018)]{gaia18} Gaia Collaboration 2018, CDS/ADC Collection of Eelectric Catalogues, 1345, 0
\bibitem[Genzel \& Downs (1977)]{gen77}  Genzel,~R., \& Downes,~D.\ 1977, \aaps, 30, 145
\bibitem[Gonidakis et~al.(2014)]{gon14} Gonidakis,~I.,  et~al.\ 2014, \mnras, 443, 3133
\bibitem[Gray(2012)]{gra12} Gray,~M.\ 2012, in {\it Maser Source in Astrophysics} (Cambridge University Press: Cambridge), Chap. 5 
\bibitem[H\"ofner et~al.(2003)]{hof03} H\"ofner,~S., et~al.\ 2003, \aap, 339, 589
\bibitem[Imai et~al.(2010)]{ima10} Imai,~H.,  et~al.\ 2010, \pasj, 62, 431
\bibitem[Imai et~al.(2003)]{ima03} Imai,~H.,  et~al.\ 2003, \apj, 590, 460
\bibitem[Imai et~al.(1997)]{ima97} Imai,~H.,  et~al.\ 1997, \aap, 317, L67
\bibitem[Kiuchi et~al.(1991)]{kiu91} Kiuchi,~H.,  et~al.\ 1991, in: Proceedings of the AGU Chapman Conference on Geodetic VLBI: 
Monitoring Global Change, NOAA Technical Report NOS 137, 35
\bibitem[Khouri et~al.(2019)]{kho19} Khouri, T.,  et~al.\ 2019, \aap, 623, L1 
\bibitem[Nakagawa et~al.(2014)]{nak14} Nakagawa,~A.,  et~al.\ 2014, PASJ, 66, 101
\bibitem[Olofsson et~al.(2002)]{olo02} 
Olofsson,~H., Gonz\'alez~Delgado,~D., Kerschbaum,~F., \&  Sch\"oier,~F.~L.\ 2002, \aap, 391, 1053
\bibitem[Ohnaka et~al.(2017)]{ohn17} Ohnaka,~K., Weigelt,~G., \& Hofmann,~K.-H.\ 2017, \aap, 597, A20 
\bibitem[Reid \& Honma(2014)]{rei14} Reid,~M.~J., \& Honma,~M.\ 2014, \araa, 52, 339 
\bibitem[Reid \& Menten(2007)]{rei07} Reid,~M.~J., \& Menten,~K.~M.\ 2007, \apj, 671, 2068
\bibitem[Reid \& Menten(1990)]{rei90} Reid,~M.~J., \& Menten,~K.~M.\ 1990, \apj, 360, L51
\bibitem[Richards et~al.(2012)]{ric12} Richards,~A.~M.~S.,  et~al.\ 2012, \aap 546, A16
\bibitem[Richards et~al.(2011)]{ric11} Richards,~A.~M.~S., Elitzur,~M., \& Yates,~J.~A.\ 2011, \aap, 525, A56
\bibitem[Richards et~al.(2010)]{ric10} Richards,~A.~M.~S.,  et~al.\ 2010, Proceedings of Science (10th EVN Symposium), 005
\bibitem[Shibata et~al.(1994)]{shi94} Shibata, K.M.,  et~al. 1994, in: VLBI TECHNOLOGY, Progress and Future Observational Possibilities, Sasao T., Manabe S., Kameya O., Inoue M. (eds.) Terra Scientific Publishing Company, p.\ 327
\bibitem[Shintani  et~al.(2008)]{shi08} Shintani,~M.,  et~al.\ 2008, \pasj, 60, 1077
\bibitem[Sudou et~al.(2002)]{sud02} Sudou,~H.,  et~al.\ 2002, \pasj, 54, 757
\bibitem[Takigawa et~al.(2017)]{tak17} Takigawa,~A.,  et~al.\ 2017, Sci. Adv., 3, 2149
\bibitem[van Leeuwen(2007)]{lee07} van~Leeuwen,~F.\ 2007, \aap, 474
\bibitem[Vlemmings et~al.(2017)]{vle17} Vlemmings,~W.,  et~al.\ 2017, Nat. Astron., 1, 848
\bibitem[Vlemmings  et~al.(2003)]{vle03}
Vlemmings,~W.~H.~T., van~Langevelde,~H.~J., Diamond,~P.~J., Habing,~H.~J., 
\& Schilizzi,~R.~T.\ 2003, \aap, 407, 213
\bibitem[Xu et~al.(2018)]{xu18} Xu,~S.,  et~al.\ 2018, \apj, 859, 14
\bibitem[Yun et~al.\ (2016)]{yun16} Yun, Y.J.,   et~al.\ 2016, \apj, 822,  3
\bibitem[Zhang et~al.(2017)]{zha17} Zhang,~B .,  et~al.\ 2017, \apj, 849, 99
\bibitem[Zhao-Geisler et~al.(2011)]{zha11}
Zhao-Geisler,~R., Quirrenbach,~A., K\"ohler,~R., Lopez,~B., \& Leinert,~C.\ 2011, \aap, 530, A120
\end{thebibliography}
\end{document}